# An Atomically Tailored Chiral Magnet with Small Skyrmions at Room Temperature

Tao Liu[1*], Camelia M. Selcu[1*], Binbin Wang[2,3], Núria Bagués[2,3], Po-Kuan Wu[1], Timothy Q. Hartnett[4], Shuyu Cheng[1], Denis Pelekhov[1], Roland A. Bennett[1], Joseph Perry Corbett[1], Jacob R. Repicky[1], Brendan McCullian[1], P. Chris Hammel[1], Jay A. Gupta[1], Mohit Randeria[1], Prasanna V. Balachandran[4,5], David W. McComb[2,3], Roland K. Kawakami[1†]

[1]Department of Physics, The Ohio State University, Columbus, Ohio 43210, USA
[2]Center for Electron Microscopy and Analysis, The Ohio State University, Columbus, Ohio 43212, USA
[3]Department of Materials Science and Engineering, The Ohio State University, OH 43210, USA
[4]Department of Materials Science and Engineering, University of Virginia, Charlottesville, VA 22902, USA
[5]Department of Mechanical and Aerospace Engineering, University of Virginia, Charlottesville, VA 22902

*These authors contributed equally to this work
†Corresponding author.    E-mail: kawakami.15@osu.edu

**Creating materials that do not exist in nature can lead to breakthroughs in science and technology. Magnetic skyrmions are topological excitations that have attracted great attention recently for their potential applications in low power, ultrahigh density memory. A major challenge has been to find materials that meet the dual requirement of small skyrmions stable at room temperature. Here we meet both these goals by developing epitaxial FeGe films with excess Fe using atomic layer molecular beam epitaxy (MBE) far from thermal equilibrium. Our novel atomic layer design permits the incorporation of 20% excess Fe while maintaining a non-centrosymmetric crystal structure supported by theoretical calculations and necessary for stabilizing skyrmions. We show that the Curie temperature is well above room temperature, and that the skyrmions probed by topological Hall effect have sizes down to 15 nm as imaged by Lorentz transmission electron microscopy (LTEM) and magnetic force microscopy (MFM). Our results illustrate new avenues for creating artificial materials tailored at the atomic scale that can impact nanotechnology.**

Magnetic skyrmions are a leading candidate for next-generation storage technology due to their potential for combining topological stability, small size, and room temperature operation along with low energies for moving or writing skyrmions[1–5]. In contrast to conventional magnetic storage where the transition from ferromagnetism to superparamagnetism limits the size reduction of data bits, topological skyrmions are not subject to this restriction so there is tremendous interest in reducing skyrmion size at room temperature. However, it is challenging to find materials that can generate skyrmions at room temperature with size below 20 nm, which is a major bottleneck for their development.[6,7] B20 materials such as FeGe and MnGe are among the most widely studied skyrmion materials[8–21] and show promising characteristics such as small size at low temperatures (~3 nm in MnGe[19,20]) and Curie temperatures above room temperature for some materials.[21,22] However, the combination of small skyrmion size with stability at room temperature remains elusive.

We start from FeGe which has a Curie temperature just below room temperature (~280 K) and skrymion size of ~80 nm.[12–18] Our strategy is to incorporate excess Fe to increase the Curie temperature while maintaining the non-centrosymmetric structure to maintain the bulk Dzyaloshinskii-Moriya interaction (DMI)[23,24] which gives rise to the twisting skyrmion spin texture (Figure 1a). To this end, we develop Fe-rich $Fe_{1.2}Ge$ using atomic layer MBE (ALMBE) where Fe and Ge atomic layers are deposited sequentially in accordance with the layered structure of FeGe(111) (Figure 1b). Under highly nonequilibrium conditions, we demonstrate the incorporation of 20% excess Fe within the B20 crystal structure to produce $Fe_{1.2}Ge$, which increases the Curie temperature to above 400 K. In terms of structure, there is no evidence of secondary phases in the transmission electron microscopy (TEM) and x-ray diffraction (XRD) results. This is supported by density functional theory (DFT) calculations showing the preferential incorporation of excess Fe atoms near specific atomic planes. Small skyrmions (<20 nm diameter) at room temperature are observed using high-resolution magnetic imaging by magnetic force microscopy (MFM) and Lorentz transmission electron microscopy (LTEM), and topological Hall resistance measurements confirm their topological nature with non-trivial winding number.

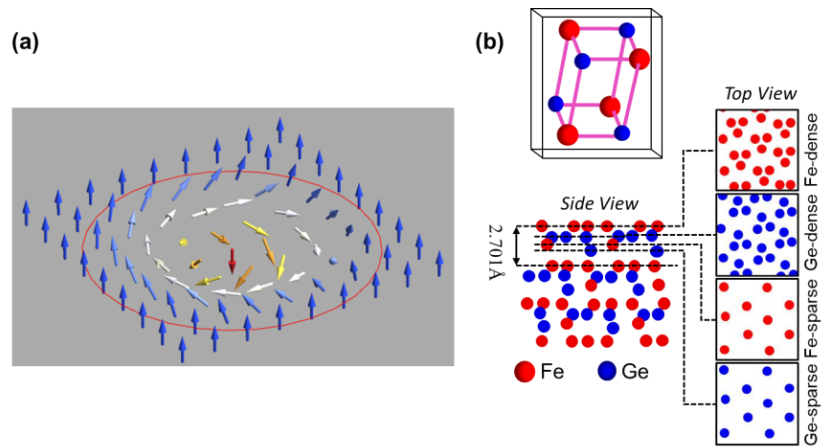

**Figure 1. Skyrmions in B20 crystals.** (a) A schematic of a localized Bloch-type skyrmion spin texture. (b) Top: the atomic unit cell of FeGe with B20 crystal structure with broken inversion symmetry. Bottom: the atomic layer structure of FeGe for (111)-oriented films, showing a sequence of dense and sparse atomic layers of Fe and Ge.

We developed Fe-rich $Fe_{1.2}Ge$ using MBE growth on top of a pure FeGe buffer layer on Si(111) (see Methods for details of MBE growth) and structural characterization by reflection high energy electron diffraction (RHEED) (Figure 2a), XRD (Figure 2b) and TEM (Figure 2c). Attempts to incorporate excess Fe atoms into FeGe using conventional MBE by co-deposition of Fe and Ge at 300 °C led to phase separation and creation of secondary phases. This can be seen by the presence of extra peaks in the XRD scan for $Fe_{1.2}Ge$ by co-deposition (green curve of Fig. 2b) and the formation of rings in the RHEED pattern (lower left image of Fig. 2a) that do not appear for pure FeGe (upper left image of Fig. 2a). Attempts to incorporate excess Fe by co-deposition within the typical growth temperature window for FeGe of 200 °C - 350 °C have failed.

As an alternative, we employed ALMBE to sequentially deposit Fe and Ge atomic layers according to the layered structure along [111] (Fig. 1b, side view) with repetitions of Ge-sparse, Fe-sparse, Ge-dense, and Fe-dense layers. As shown in Fig. 1b (top view), the dense layers have a three atom basis on a hexagonal lattice and the sparse layers have a one atom basis on a hexagonal lattice[25]. The growth of stoichiometric FeGe by ALMBE proceeded successfully over an expanded growth temperature window down to room temperature (RT), where the RHEED pattern is streaky (upper right image of Fig. 2a) but slightly wider than for conventional growth (upper left image of Fig. 2a). Under these highly nonequilibrium conditions, we grew Fe-rich $Fe_{1.2}Ge$ using RT-ALMBE by increasing the growth time of the Fe-sparse layer by 80%. We observed no discernible changes in the RHEED pattern (lower right image of Fig. 2a) compared to stoichiometric FeGe and no obvious signs of secondary phases appeared. This is further supported by XRD scans (red curve, Fig. 2b) which show no additional peaks compared to pure FeGe (blue curve, Fig. 2b). The one notable difference is that the B20 FeGe peak has a slight shift to lower angles (Figure 2b inset) which indicates that the Fe-rich $Fe_{1.2}Ge$ has a 0.5% expanded lattice parameter along the growth direction.

The atomic-scale structure of Fe-rich $Fe_{1.2}Ge$ was characterized by cross-sectional scanning (S)TEM.

Figure 2c shows a STEM high angle annular dark-field (HAADF) image of the interfacial region between the Fe$_{1.2}$Ge film and the FeGe buffer. Most notable is the high quality of the Fe$_{1.2}$Ge layer, which appears to be very similar to the FeGe buffer layer (grown at 300 °C). HAADF-STEM images also show the Fe-rich $Fe_{1.2}Ge$ film has grains of various sizes, from a few tens to few hundreds of nanometers, which reflects the size and orientation of the grains formed in the FeGe buffer layer[26,27]. Although the presence of the excess Fe is not apparent in the HAADF image, measurement of the composition using x-ray energy dispersive spectroscopy (XEDS) shows that the average Fe content in the Fe-rich $Fe_{1.2}Ge$ is 18% higher than in the FeGe buffer (Figure 2d). Regarding the possible presence of secondary phases, neither the HAADF image nor the XEDS image provides any evidence for the formation of secondary phases or compositional inhomogeneity. Interestingly, the excess Fe is not resolved in the Fe rich region in the atomic resolution HAADF image, suggesting that they may have randomly occupied sites. This is supported by DFT calculations discussed next.

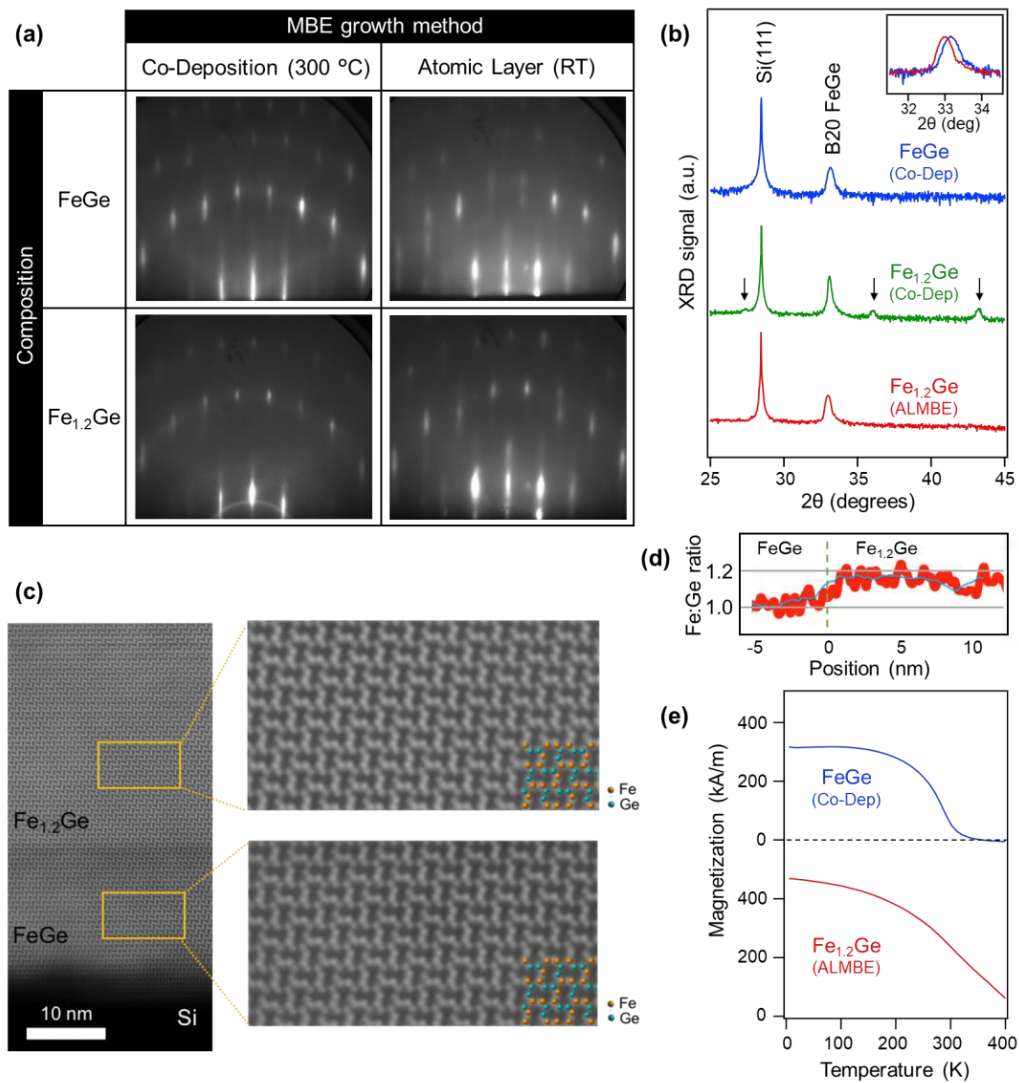

**Figure 2. Growth, structure, and magnetic properties of Fe-rich Fe$_{1.2}$Ge.** (a) RHEED patterns for conventional FeGe co-deposited at 300 °C, Fe$_{1.2}$Ge co-deposited at 300 °C, and FeGe and Fe$_{1.2}$Ge grown by atomic layer MBE (ALMBE) at room temperature. (b) θ-2θ x-ray diffraction scans for co-deposited FeGe (blue), co-deposited Fe$_{1.2}$Ge (green), and ALMBE Fe$_{1.2}$Ge (red) show the presence of secondary phases (black arrows) for co-deposited Fe$_{1.2}$Ge but not for ALMBE Fe$_{1.2}$Ge. Inset: A close-up of the B20 FeGe peak for normal co-deposited FeGe (blue) and ALMBE Fe$_{1.2}$Ge (red) indicates an expansion of the B20 lattice due to the incorporation of excess Fe. (c) Cross-sectional HAADF-STEM image of Fe$_{1.2}$Ge on FeGe/Si(111) showing the B20 lattice. (d) XEDS scan showing a higher Fe:Ge ratio in the Fe$_{1.2}$Ge film. (e) Magnetization as a function of temperature with out-of-plane field of 2 Tesla for FeGe (blue) and Fe-rich Fe$_{1.2}$Ge (red).

The ability to incorporate excess Fe atoms into the FeGe film was studied theoretically using DFT calculations of bulk FeGe with additional Fe atoms in the unit cell, where energy minimization was performed while allowing the lattice parameter along one of the [111] directions to vary (see Methods). With 12 Fe and 12 Ge atoms in the unit cell for the calculation, the addition of 1, 2, or 3 Fe atoms corresponds to an average excess Fe concentration of 8.3% ($Fe_{1.08}Ge$), 16.7% ($Fe_{1.17}Ge$), or 25.0% ($Fe_{1.25}Ge$), respectively. In each case, the calculation successfully converged to a local energy minimum, and the excess Fe atom(s) were located near Fe-sparse layers. This supports our ALMBE design to deposit excess Fe into the Fe-sparse layers. Regarding the HAADF image, since there are multiple sites for excess Fe and the site-selection is subject to randomness (layer and position within layer), the imaging through the cross-section (e.g. an atom in the HAADF image is generated by a column of aligned atoms) combined with randomness causes the excess Fe atoms to average away. An extended discussion of the DFT results are included in the Supplementary Information (SI) section 1.

Magnetic properties investigated by SQUID magnetometry show an enhanced temperature dependence for Fe-rich $Fe_{1.2}Ge$ compared to conventional FeGe. Applying a 2 Tesla out-of-plane field to remove any spin texture, the temperature scan shows that the field-polarized magnetization persists to above 400 K for Fe-rich $Fe_{1.2}Ge$ (Figure 2e, red curve) while the magnetization disappears at ~300 K for FeGe (Figure 2e, blue curve).

Direct imaging of room temperature skyrmions in Fe-rich $Fe_{1.2}Ge$ was realized by MFM with a high-resolution tip and LTEM. Figure 3 shows room temperature MFM images obtained from a 108 nm thick Fe-rich $Fe_{1.2}Ge$ film for different out-of-plane fields (see Methods for details of the measurement). At low magnetic fields of 0 mT and 50 mT (Figures 3a and 3b), the magnetic domain structure consists of disordered stripes in white (i.e. helical phase) and a few white dots (i.e. skyrmions). As the magnetic field is increased to 150 mT (Figure 3c), the stripes disappear and give way to the formation of many discontinuous dots. With further increasing field (Figures 3d, 3e, 3f), the dots gradually disappear until they are virtually gone at 300 mT. This trend of magnetic domain structure as a function of magnetic field is very similar to what has been observed in FeGe and other B20 skyrmion materials[12,13,16,18,28]. Moreover, the MFM data agrees with the topological Hall resistivity (THR) shown in Figure 3g as a function of out-of-plane magnetic field and temperature, obtained following the procedure outlined in SI section 3. Notably, skyrmions generate THR but helices do not. The general features of the THR phase diagram for Fe-rich $Fe_{1.2}Ge$ (Fig. 3g) and pure FeGe (Fig. 3h) are similar with THR exhibiting similar magnitudes, maxima occurring at finite field, and a reduction of signal at low temperatures. Interestingly, the THR for Fe-rich $Fe_{1.2}Ge$ extends to much higher temperatures with substantial magnitude even at 340 K, which is the high temperature limit of our instrument. Comparing the THR for $Fe_{1.2}Ge$ with the MFM images reveals a good agreement. At room temperature, the THR exhibits a maxima within 150 – 180 mT, where the MFM shows a high density of dots. Increasing the field above 150 mT, the dots gradually disappear and are nearly gone by 300 mT, which also agrees with the THR. These observations confirm that the localized dot-like magnetic textures are indeed skyrmions, which have a non-zero topological winding number to generate THR. We estimate the magnitude of THR using a Drude model analysis[29] and obtain values within a factor of 2-3 of the experimental values (see SI section 4).

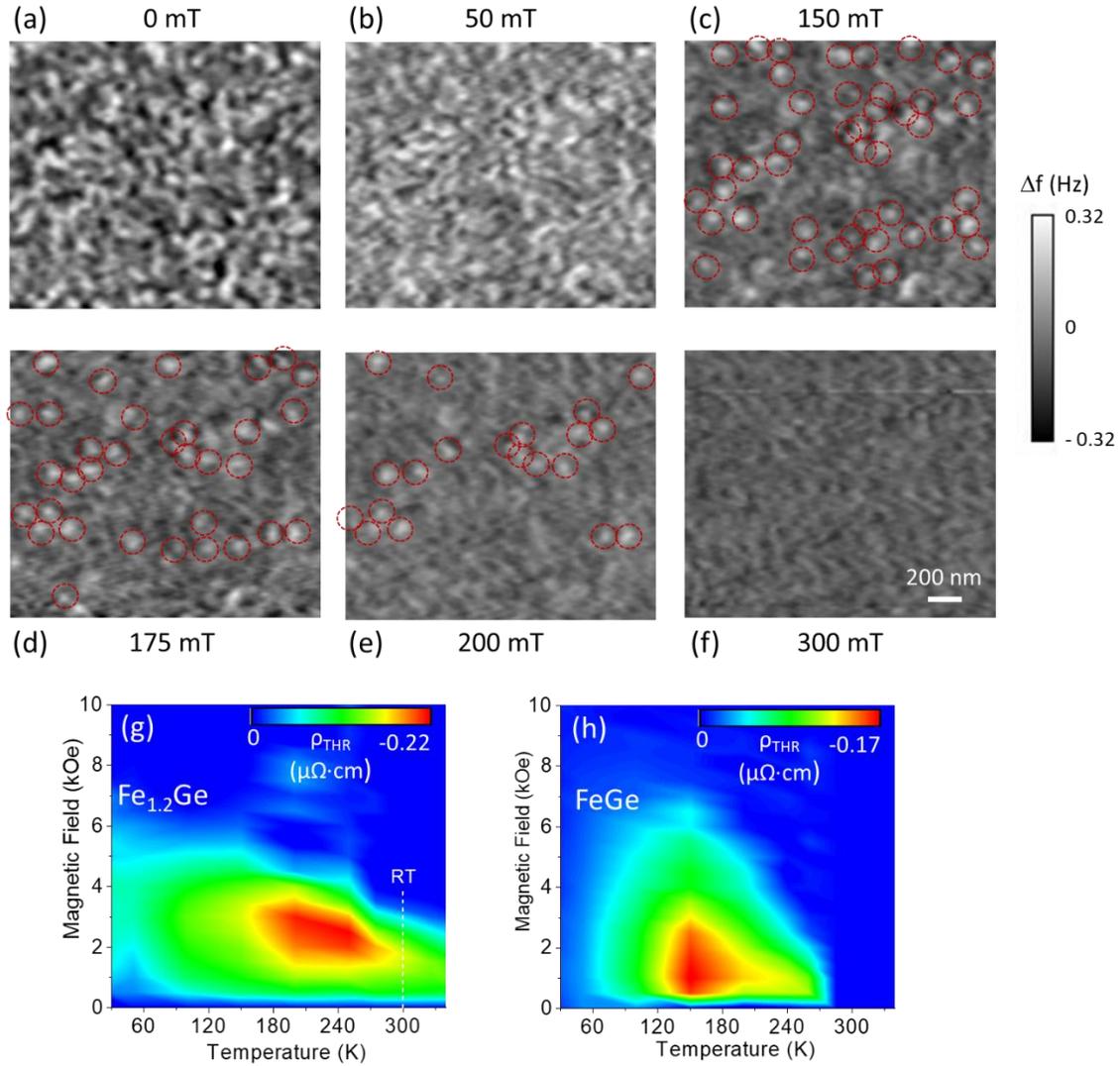

**Figure 3. MFM imaging and topological Hall effect.** Room temperature MFM images of a 108 nm $Fe_{1.2}Ge$ film, with out-of-plane magnetic fields of (a) 0 mT, (b) 50 mT, (c) 150 mT, (d) 175 mT, (e) 200 mT, (f) 300 mT. Topological Hall resistivity for (g) Fe-rich $Fe_{1.2}Ge$ and (h) FeGe. MFM images without circles is provided in SI section 2.

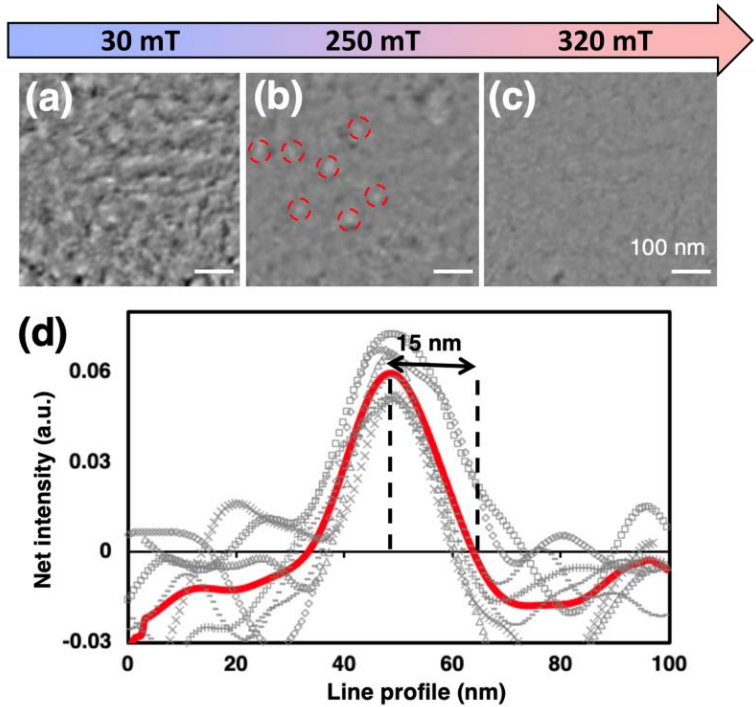

**Figure 4. Lorentz TEM imaging.** Plan-view LTEM images, after removing the background, of a 108 nm $Fe_{1.2}Ge$ film acquired at room temperature with out-of-plane magnetic fields: (a) 30 mT, (b) 250 mT, and (c) 320 mT. (d) The line profiles of the magnetic skyrmions highlighted with red circles in (b). The red curve is an average of the line profiles.

To determine the skyrmion sizes, we studied $Fe_{1.2}Ge$ films using LTEM in the plan-view. The LTEM raw images are strongly influenced by diffraction contrast (e. g. from grain boundaries). In order to enhance the magnetic contrast, as shown in Figure 4a-c, we use a background subtraction procedure described in the Methods. The change in magnetic contrast in LTEM as a function of applied field is consistent with the MFM results. At low magnetic fields, magnetic textures are stronger and randomly distributed as stripes and dots (Figure 4a); while most of them fade off with the increase of magnetic fields, as shown in Figure 4b and 4c. In LTEM, Bloch type skyrmions appear as either bright or dark dots depending on ther chirality.[27,30] To measure the size of the skyrmions, we averaged the bright dots in Figure 4b that clearly disappeared in Figure 4c. The averaged skyrmion size is identified to be ~ 15 nm at room temperature.

Finally, we discuss the micromagnetic parameters leading to the spin texture: the saturation magnetization $M_s$, exchange stiffness $A_{ex}$, bulk Dzyaloshinskii-Moriya interaction $D$, magnetic anisotropy $K_u$, and the Gilbert damping $\alpha$. One of the key questions in skyrmion materials is: what parameters determine the size and stability of skyrmions? The simplest answer is that the size of skyrmions in determined by the ratio of the exchange to DMI, namely by the length scale $L_D = 4\pi A_{ex}/D$ [31,32]. However, there are several theoretical studies, especially of isolated skyrmions, which conclude that their size is *not* determined by $L_D$, and that it depends in an essential way on $K_u$ and $M_s$ [33–35]. In view of this, it is especially important to understand the parameters that characterize small, room-temperature skyrmions in Fe-rich FeGe. As discussed in detail in SI section 2, the combination of micromagnetic simulations with our experimental measurements (magnetization, ferromagnetic resonance, MFM and LTEM) show that the size of Bloch skyrmions is indeed determined by $L_D \sim A_{ex}/D$ and weakly dependent of $K_u$ and $M_s$. Quantitative estimates of the micromagnetic parameters are provided in SI section 5.

In conclusion, the highly-controlled nonequilibrium synthesis of Fe-rich $Fe_{1.2}Ge$ by RT-ALMBE has produced small skyrmions at room temperature in an epitaxial film. The incorporation of excess Fe into FeGe pushes the Curie temperature over room temperature, and MFM, LTEM, and THR identify room temperature skyrmions with a mean diameter as small as ~15 nm. The realization of small skyrmions at room temperature is significant for the development of ultrahigh density, low power magnetic memories, and the ability to super-saturate the B20 lattice using nonequilibrium RT-ALMBE opens new avenues for atomic-scale materials design.

**Methods**

**Epitaxial growth:**
Epitaxial FeGe and $Fe_{1.2}Ge$ films were grown on Si(111) substrates in an MBE chamber with a base pressure of $2\times10^{-10}$ Torr. Si(111) substrates were prepared by sonicating in acetone for 5 min and isopropyl alcohol for 5 min. The cleaned wafers were dipped in buffered HF for 2 min, to remove the native oxide and terminate the dangling Si bonds with H, then rinsed in DI water. The substrates were immediately mounted onto Ta flag-style paddles, inserted into the growth chamber, and annealed at 800 °C for 20 min to desorb the hydrogen to obtain a 7×7 reconstruction. Thermal effusion cells were used for the deposition of Fe and Ge with the typical elemental growth rates in the range of 1-4 Å/min. *In situ* RHEED was used to monitor the crystallographic structure and atomic-scale roughness of the surface during growth. All samples began with a buffer layer of FeGe grown at 300 °C, using co-deposition of Fe and Ge flux matched with a 1:1 ratio.

**Magnetic force microscopy:**
MFM measurements were performed in a Quantum Design PPMS system to provide variable magnetic fields and temperatures. An MFM insert and high resolution tips were used to probe the Fe-rich FeGe films. The MFM images were acquired at constant height mode for a series of magnetic fields. In order to obtain the magnetic contribution to the frequency shifts, we subtracted the background signal[36] obtained by imaging at a high magnetic field of 1.5 T where the sample is magnetically saturated.

**Transmission electron microscopy:**
The TEM investigations were carried out using a Thermo Scientific Image TITAN G2 60-300 microscope equipped with a Lorentz lens and image corrector, operated at 300 kV. The plan view TEM specimens were prepared from a $Fe_{1.2}Ge$ (108 nm) thin film on FeGe buffer grown on Si(111) substrate using a wedge mechanical polishing method. The cross-section TEM specimens for HAADF-STEM characterization and XEDS analyses were prepared by Ga focused ion beam (FIB). The STEM XEDS spectrum-imaging was performed using quad-silicon drift detectors, and the line-profile of elements ratio was measured along with the sample grown direction. Element ratio was normalized using the FeGe buffer layer as a reference, assuming Fe/Ge = 1.

LTEM images were taken with a defocus of 100 μm to observe the magnetic contrast. The applied vertical magnetic field was controlled by varying the objective lens current of TEM. The actual magnetic fields were calibrated by a home-built TEM holder with a Hall-effect sensor. In our thin films, strong diffraction contrast is observed in LTEM mode, which reduces the visibility of magnetic contrast contributions. In order to reduce the diffraction contrast contributions in the LTEM image, we first recorded a series of LTEM images from the same region of interest at different magnetic fields. Then, the LTEM image recorded at 3600 Oe, considering the sample is saturated and no magnetic contrast is present at that field, was subtracted as "background" contrast from all other images after some careful alignments and intensity normalization. More details of a similar procedure to enhance the magnetic contrast could be found from our earlier reports in a FeGe thin film plan view specimen[26]. In addition, Gaussian blurring of 4 pixel standard deviation (~ 6 nm) was performed to remove high-frequency noise in the background-removed LTEM images. The skyrmion size is measured from the FWHM of a line profile.

**Density functional theory:**

*Ab initio* electronic structure calculations were carried out in the Density Functional Theory (DFT) framework using the planewave pseudopotential Quantum ESPRESSO code[37]. Core and valence electrons were treated using the ultrasoft pseudopotential method[38]. The exchange-correlation functionals were described using the Perdew-Burke-Ernzerhof parameterization of the generalized gradient approximation modified for solids (PBEsol[39]). To emulate the effects of epitaxial thin film growth the structure was oriented in a (111)-crystallographic plane. In addition, bulk structures were also considered for comparative studies. In the case of simulations involving (111)-oriented FeGe, excess Fe-atoms were added to the structure, which were then allowed to relax by expanding the unit cell only in the z-direction until an energy convergence threshold of $10^{-8}$ eV and Hellmann-Feynman forces less than 2 meV/Å were achieved. The plane wave cutoff energy was set to 60 Ry, the charge density cutoff was set at 600 Ry, and a Γ-centered Monkhorst-Pack *k*-point mesh with dimensions 6x6x6 was used to sample the Brillouin Zone. The Brillouin zone integration used Marzari-Vanderbilt smearing[40] with a smearing width of 0.27 eV. The scalar relativistic pseudopotentials were taken from the PSLibrary[41]. A collinear ferromagnetic spin order was imposed on all the Fe-atoms (including the interstitial Fe-atoms). Regardless of the initial position of the interstitial Fe-atoms, all structures relaxed to form a new Fe sparse layer in the structure. The converged out-of-plane lattice constant is 9.5369 Å.

**Acknowledgements**
This work was supported by the DARPA TEE program under Grants Nos. D18AP00008 and D18AP00009. This research was partially supported by the Center for Emergent Materials, an NSF MRSEC, under award number DMR-2011876.

**Author contributions**
TL, RAB, SC, and RKK performed the MBE growth, SQUID, XRD, and THR measurements. CMS, JPC, JRR, and JAG performed the MFM measurements and analysis. BW, NB, and DWM performed the STEM and LTEM measurements and analysis. TQH, PVB, DP, and MR performed the theoretical analysis. BM and PCH performed the FMR measurements. All authors contributed to the preparation of the manuscript.

**Competing financial interests**

The authors declare no competing financial interests.

**References**

1. Fert, A., Cros, V. & Sampaio, J. Skyrmions on the track. *Nature Nanotechnology* **8,** 152–156 (2013).

2. Wiesendanger, R. Nanoscale magnetic skyrmions in metallic films and multilayers: a new twist for spintronics. *Nature Reviews Materials* **1,** 1–11 (2016).

3. Fert, A., Reyren, N. & Cros, V. Magnetic skyrmions: advances in physics and potential applications. *Nature Reviews Materials* **2,** 1–15 (2017).

4. Jiang, W., Chen, G., Liu, K., Zang, J., te Velthuis, S. G. E. & Hoffmann, A. Skyrmions in magnetic multilayers. *Physics Reports* **704,** 1–49 (2017).

5. Back, C., Cros, V., Ebert, H., Everschor-Sitte, K., Fert, A., Garst, M., Ma, T., Mankovsky, S., Monchesky, T. L., Mostovoy, M., Nagaosa, N., Parkin, S. S. P., Pfleiderer, C., Reyren, N., Rosch, A., Taguchi, Y., Tokura, Y., Bergmann, K. von & Zang, J. The 2020 skyrmionics roadmap. *J. Phys. D: Appl. Phys.* **53,** 363001 (2020).

6. Everschor-Sitte, K., Masell, J., Reeve, R. M. & Kläui, M. Perspective: Magnetic skyrmions— Overview of recent progress in an active research field. *Journal of Applied Physics* **124,** 240901 (2018).

7. Caretta, L., Mann, M., Büttner, F., Ueda, K., Pfau, B., Günther, C. M., Hessing, P., Churikova, A., Klose, C., Schneider, M., Engel, D., Marcus, C., Bono, D., Bagschik, K., Eisebitt, S. & Beach, G. S. D. Fast current-driven domain walls and small skyrmions in a compensated ferrimagnet. *Nature Nanotechnology* **13,** 1154–1160 (2018).

8. Mühlbauer, S., Binz, B., Jonietz, F., Pfleiderer, C., Rosch, A., Neubauer, A., Georgii, R. & Böni, P. Skyrmion Lattice in a Chiral Magnet. *Science* **323,** 915–919 (2009).

9. Adams, T., Mühlbauer, S., Pfleiderer, C., Jonietz, F., Bauer, A., Neubauer, A., Georgii, R., Böni, P., Keiderling, U., Everschor, K., Garst, M. & Rosch, A. Long-Range Crystalline Nature of the Skyrmion Lattice in MnSi. *Phys. Rev. Lett.* **107,** 217206 (2011).

10. Nii, Y., Nakajima, T., Kikkawa, A., Yamasaki, Y., Ohishi, K., Suzuki, J., Taguchi, Y., Arima, T., Tokura, Y. & Iwasa, Y. Uniaxial stress control of skyrmion phase. *Nature Communications* **6,** 8539 (2015).

11. Li, Y., Kanazawa, N., Yu, X. Z., Tsukazaki, A., Kawasaki, M., Ichikawa, M., Jin, X. F., Kagawa, F. & Tokura, Y. Robust Formation of Skyrmions and Topological Hall Effect Anomaly in Epitaxial Thin Films of MnSi. *Phys. Rev. Lett.* **110,** 117202 (2013).

12. Yu, X. Z., Kanazawa, N., Onose, Y., Kimoto, K., Zhang, W. Z., Ishiwata, S., Matsui, Y. & Tokura, Y.

Near room-temperature formation of a skyrmion crystal in thin-films of the helimagnet FeGe. *Nature Materials* **10,** 106–109 (2011).

13. Huang, S. X. & Chien, C. L. Extended Skyrmion Phase in Epitaxial FeGe(111) Thin Films. *Phys. Rev. Lett.* **108,** 267201 (2012).

14. Zhang, L., Han, H., Ge, M., Du, H., Jin, C., Wei, W., Fan, J., Zhang, C., Pi, L. & Zhang, Y. Critical phenomenon of the near room temperature skyrmion material FeGe. *Sci Rep* **6,** 22397 (2016).

15. Kovács, A., Caron, J., Savchenko, A. S., Kiselev, N. S., Shibata, K., Li, Z.-A., Kanazawa, N., Tokura, Y., Blügel, S. & Dunin-Borkowski, R. E. Mapping the magnetization fine structure of a lattice of Bloch-type skyrmions in an FeGe thin film. *Appl. Phys. Lett.* **111,** 192410 (2017).

16. Gallagher, J. C., Meng, K. Y., Brangham, J. T., Wang, H. L., Esser, B. D., McComb, D. W. & Yang, F. Y. Robust Zero-Field Skyrmion Formation in FeGe Epitaxial Thin Films. *Phys. Rev. Lett.* **118,** 027201 (2017).

17. Leroux, M., Stolt, M. J., Jin, S., Pete, D. V., Reichhardt, C. & Maiorov, B. Skyrmion Lattice Topological Hall Effect near Room Temperature. *Sci Rep* **8,** 15510 (2018).

18. Ahmed, A. S., Rowland, J., Esser, B. D., Dunsiger, S. R., McComb, D. W., Randeria, M. & Kawakami, R. K. Chiral bobbers and skyrmions in epitaxial FeGe/Si(111) films. *Phys. Rev. Materials* **2,** 041401(R) (2018).

19. Kanazawa, N., Onose, Y., Arima, T., Okuyama, D., Ohoyama, K., Wakimoto, S., Kakurai, K., Ishiwata, S. & Tokura, Y. Large Topological Hall Effect in a Short-Period Helimagnet MnGe. *Phys. Rev. Lett.* **106,** 156603 (2011).

20. Tanigaki, T., Shibata, K., Kanazawa, N., Yu, X., Onose, Y., Park, H. S., Shindo, D. & Tokura, Y. Real-Space Observation of Short-Period Cubic Lattice of Skyrmions in MnGe. *Nano Lett.* **15,** 5438–5442 (2015).

21. Balasubramanian, B., Manchanda, P., Pahari, R., Chen, Z., Zhang, W., Valloppilly, S. R., Li, X., Sarella, A., Yue, L., Ullah, A., Dev, P., Muller, D. A., Skomski, R., Hadjipanayis, G. C. & Sellmyer, D. J. Chiral Magnetism and High-Temperature Skyrmions in B20-Ordered Co-Si. *Phys. Rev. Lett.* **124,** 057201 (2020).

22. Budhathoki, S., Sapkota, A., Law, K. M., Ranjit, S., Nepal, B., Hoskins, B. D., Thind, A. S., Borisevich, A. Y., Jamer, M. E., Anderson, T. J., Koehler, A. D., Hobart, K. D., Stephen, G. M., Heiman, D., Mewes, T., Mishra, R., Gallagher, J. C. & Hauser, A. J. Room-temperature skyrmions in strain-engineered FeGe thin films. *Phys. Rev. B* **101,** 220405 (2020).

23. Dzyaloshinsky, I. A thermodynamic theory of "weak" ferromagnetism of antiferromagnetics. *Journal of Physics and Chemistry of Solids* **4,** 241–255 (1958).

24. Moriya, T. Anisotropic Superexchange Interaction and Weak Ferromagnetism. *Phys. Rev.* **120,** 91–98 (1960).

25. Corbett, J. P., Zhu, T., Ahmed, A. S., Tjung, S. J., Repicky, J. J., Takeuchi, T., Guerrero-Sanchez, J., Takeuchi, N., Kawakami, R. K. & Gupta, J. A. Determining Surface Terminations and Chirality of Noncentrosymmetric FeGe Thin Films via Scanning Tunneling Microscopy. *ACS Appl. Mater. Interfaces* **12,** 9896–9901 (2020).

26. Bagues, N., Wang, B., Liu, T., Selcu, C., Boona, S., Kawakami, R., Randeria, M. & McComb, D. Imaging of Magnetic Textures in Polycrystalline FeGe Thin Films via in-situ Lorentz Transmission Electron Microscopy. *Microscopy and Microanalysis* **26,** 1700–1702 (2020).

27. Wang, B., Bagués, N., Liu, T., Kawakami, R. K. & McComb, D. W. Extracting weak magnetic contrast from complex background contrast in plan-view FeGe thin films. *Ultramicroscopy* **232,** 113395 (2022).

28. Kanazawa, N., Shibata, K. & Tokura, Y. Variation of spin–orbit coupling and related properties in skyrmionic system $Mn_{1-x}Fe_xGe$. *New J. Phys.* **18,** 045006 (2016).

29. Bruno, P., Dugaev, V. K. & Taillefumier, M. Topological Hall Effect and Berry Phase in Magnetic Nanostructures. *Phys. Rev. Lett.* **93,** 096806 (2004).

30. Nguyen, K. X., Zhang, X. S., Turgut, E., Cao, M. C., Glaser, J., Chen, Z., Stolt, M. J., Chang, C. S., Shao, Y.-T., Jin, S., Fuchs, G. D. & Muller, D. A. Disentangling Magnetic and Grain Contrast in Polycrystalline $\mathrm{Fe}\mathrm{Ge}$ Thin Films Using Four-Dimensional Lorentz Scanning Transmission Electron Microscopy. *Phys. Rev. Appl.* **17,** 034066 (2022).

31. Nagaosa, N. & Tokura, Y. Topological properties and dynamics of magnetic skyrmions. *Nature Nanotechnology* **8,** 899–911 (2013).

32. Banerjee, S., Rowland, J., Erten, O. & Randeria, M. Enhanced Stability of Skyrmions in Two-Dimensional Chiral Magnets with Rashba Spin-Orbit Coupling. *Phys. Rev. X* **4,** 031045 (2014).

33. Büttner, F., Lemesh, I. & Beach, G. S. D. Theory of isolated magnetic skyrmions: From fundamentals to room temperature applications. *Sci. Rep.* **8,** 4464 (2018).

34. Wang, X. S., Yuan, H. Y. & Wang, X. R. A theory on skyrmion size. *Commun. Phys.* **1,** 1–7 (2018).

35. Bera, S. & Mandal, S. S. Theory of the skyrmion, meron, antiskyrmion, and antimeron in chiral magnets. *Phys. Rev. Research* **1,** 033109 (2019).

36. Meng, K.-Y., Ahmed, A. S., Baćani, M., Mandru, A.-O., Zhao, X., Bagués, N., Esser, B. D., Flores, J., McComb, D. W., Hug, H. J. & Yang, F. Observation of Nanoscale Skyrmions in $SrIrO_3$/$SrRuO_3$ Bilayers. *Nano Lett.* **19,** 3169–3175 (2019).

37. Giannozzi, P., Baroni, S., Bonini, N., Calandra, M., Car, R., Cavazzoni, C., Ceresoli, D., Chiarotti, G. L., Cococcioni, M., Dabo, I., Corso, A. D., Gironcoli, S. de, Fabris, S., Fratesi, G., Gebauer, R., Gerstmann, U., Gougoussis, C., Kokalj, A., Lazzeri, M., Martin-Samos, L., Marzari, N., Mauri, F., Mazzarello, R., Paolini, S., Pasquarello, A., Paulatto, L., Sbraccia, C., Scandolo, S., Sclauzero, G., Seitsonen, A. P., Smogunov, A., Umari, P. & Wentzcovitch, R. M. QUANTUM ESPRESSO: a modular and open-source software project for quantum simulations of materials. *J. Phys.: Condens. Matter* **21,** 395502 (2009).

38. Vanderbilt, D. Soft self-consistent pseudopotentials in a generalized eigenvalue formalism. *Phys. Rev. B* **41,** 7892–7895 (1990).

39. Perdew, J. P., Ruzsinszky, A., Csonka, G. I., Vydrov, O. A., Scuseria, G. E., Constantin, L. A., Zhou, X. & Burke, K. Restoring the Density-Gradient Expansion for Exchange in Solids and Surfaces. *Phys. Rev. Lett.* **100,** 136406 (2008).

40. Marzari, N., Vanderbilt, D., De Vita, A. & Payne, M. C. Thermal Contraction and Disordering of the Al(110) Surface. *Phys. Rev. Lett.* **82,** 3296–3299 (1999).

41. Dal Corso, A. Pseudopotentials periodic table: From H to Pu. *Computational Materials Science* **95,** 337–350 (2014).

# Supplementary Information:

## An Atomically Tailored Chiral Magnet for Small Skyrmions at Room Temperature

Tao Liu[1*], Camelia M. Selcu[1*], Binbin Wang[2,3], Núria Bagués[2,3], Po-Kuan Wu[1], Timothy Q. Hartnett[4], Shuyu Cheng[1], Denis Pelekhov[1], Roland A. Bennett[1], Joseph Perry Corbett[1], Jacob R. Repicky[1], Brendan McCullian[1], P. Chris Hammel[1], Jay A. Gupta[1], Mohit Randeria[1], Prasanna V. Balachandran[4,5], David W. McComb[2,3], Roland K. Kawakami[1†]

[1]Department of Physics, The Ohio State University, Columbus, Ohio 43210, USA
[2]Center for Electron Microscopy and Analysis, The Ohio State University, Columbus, Ohio 43212, USA
[3]Department of Materials Science and Engineering, The Ohio State University, OH 43210, USA
[4]Department of Materials Science and Engineering, University of Virginia, Charlottesville, VA 22902, USA
[5]Department of Mechanical and Aerospace Engineering, University of Virginia, Charlottesville, VA 22902

*These authors contributed equally to this work
†Corresponding author. E-mail: kawakami.15@osu.edu

**Contents**

**1. Density Functional Theory Calculations of Fe-rich FeGe**

**2. MFM Images Without Circles**

**3. Topological Hall Effect Measurements and Analysis**

**4. Topological Hall Resistivity (THR) Estimates**

**5. Determination of Micromagnetic Parameters**

# 1. Density Functional Theory Calculations of Fe-rich FeGe

## Overview

A series of density functional theory (DFT) calculations present a consistent picture indicating the preferential occupation of excess Fe atoms in the sparse layers of the B20 crystal. These calculations do not attempt to model the complex growth kinetics and do not consider interfaces with substrates. Instead, the DFT calculations are bulk-like with unit cells and periodic boundary conditions. To connect the bulk-like calculations to (111) films, the in-plane lattice parameters within the (111) planes are held fixed while the out-of-plane lattice parameter between the (111) planes is allowed to relax. The energetics of Fe incorporation are investigated by adding excess Fe atoms and allowing the system to converge to a minimum energy configuration. While this does not model the complexities of real thin film systems, it nevertheless illustrates the preferential incorporation of excess Fe into the sparse layers, consistent with the experimental protocol for atomic layer MBE.

## Supercells of (111)-oriented FeGe

The unit cell of a bulk FeGe contains eight atoms (4 Fe and 4 Ge). All Fe- and Ge-atoms share a unique Wyckoff site each. All DFT calculations were performed using a (111)-oriented FeGe supercell that contained 24 atoms (12 Fe and 12 Ge). See Methods section in the main manuscript for additional information. The DFT optimized in-plane and out-of-plane lattice constants for the stoichiometric (111)-FeGe are 6.5053 and 7.9673 Å, respectively. The crystal structure is shown in **Figure S1**. The dense and sparse atomic arrangements of Fe atoms in the structure are also labeled in **Figure S1**. This specific supercell configuration support three dense and sparse layers of Fe atoms. We used this stoichiometric atomic arrangement as a baseline, and systematically introduced excess Fe atoms into the structure to understand its preferential occupation in the interstitial sites. We investigated three additional non-stoichiometric compositions using DFT calculations: $Fe_{1.08}Ge$ (with 1 excess Fe atom), $Fe_{1.17}Ge$ (with 2 excess Fe atoms), and $Fe_{1.25}Ge$ (with 3 excess Fe atoms).

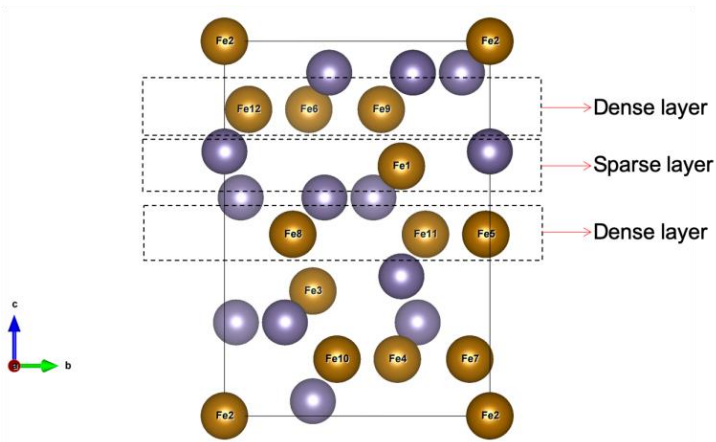

**Figure S1**. Crystal structure of a bulk (111)-oriented stoichiometric FeGe. The dense and sparse arrangements of Fe atoms are labeled. The brown and purple atoms correspond to Fe and Ge, respectively.

## $Fe_{1.08}Ge$ with one excess Fe atom:

Our first set of calculations involved introducing one excess Fe atom in the dense and sparse layers. In our calculations, the in-plane lattice constants were fixed, whereas the out-of-plane lattice constants were allowed to relax. All internal coordinates were allowed to relax. The fully relaxed crystal structures are shown in **Figure S2**. We were able to identify a local minimum for the excess Fe atom in both the structures (labeled as "Fe13" in **Figure S2**). In **Figure S2a**, the excess Fe atom occupies the dense layer and in **Figure S2b**, it occupies the sparse layer. Further, the crystal structure shown in **Figure S2b,** where we have an excess Fe atom in the sparse layer, was 20 meV/atom *lower* in total energy compared to the structure shown in **Figure S2a**. This initial result indicated the preferential occupation of the excess Fe atom in the sparse layers of the (111)-oriented FeGe structure. The out-of-plane lattice constant increased to 8.2636 Å for the lowest energy structure (**Figure S2b**).

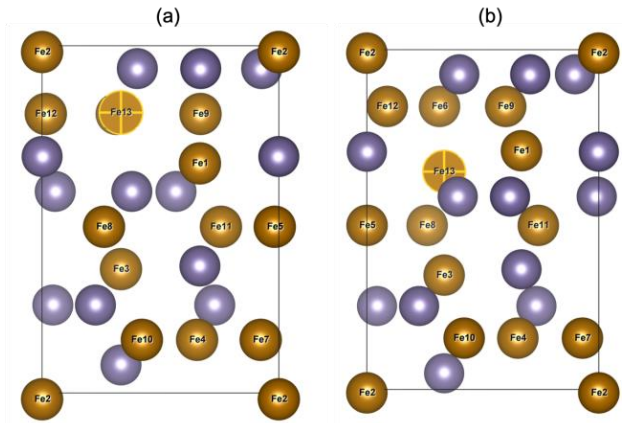

**Figure S2**. Crystal structures of fully relaxed $Fe_{1.08}Ge$ composition with one excess Fe atom (labeled as "Fe13") in the (a) Dense and (b) Sparse layers.

## $Fe_{1.17}Ge$ with two excess Fe atoms:

Our second set of calculations involved introducing two excess Fe atoms into the structure. The composition of the new structure corresponds to $Fe_{1.17}Ge$, which is closer to that of the experimental $Fe_{1.2}Ge$ composition. Informed by the first set of results from DFT, we introduced the two excess Fe atoms into the sparse layers. Our choice of the (111)-FeGe supercell permitted three different locations for the distribution of the excess two Fe atoms in the three sparse layers. We explored all three locations. Similar to the previous calculation, the in-plane lattice constants were fixed and the out-of-plane lattice constants were allowed to relax. All internal coordinates were also allowed to relax. The results from the converged DFT calculations are shown in **Figure S3**. All three configurations have unique local minima and the excess Fe atoms occupy the sparse layers. Intriguingly, the excess Fe atoms form their own *new* sparse layers in the structure. All three structures were practically degenerate (the energy difference was of the order of 0.01 meV/atom). The out-of-plane lattice constant increased to 8.903-8.904 Å.

In addition to the configurations shown in **Figure S3**, we also tested another atomic arrangement. In this case, we initialized the two excess Fe atoms in the dense layers and let all atoms to relax to equilibrium positions. The out-of-plane lattice constant was also allowed to relax. The initial and fully relax crystal structures are shown in **Figure S4a** and **Figure S4b**, respectively. The DFT calculations were initialized with both excess Fe atoms co-planar in the dense layer (the *z*-coordinate of the excess and native Fe atoms in the dense layers are shown in **Figure S4a**). In the relaxed configuration, there were two interesting outcomes. First, the excess atom Fe13 diffused out of the dense layer and formed its own sparse layer. This atomic arrangement is consistent with the expected trend seen in **Figures S2** and S**3**. Second, the excess atom Fe14 remained in close proximity to the initial position in the dense layer, but disrupted the symmetry and coplanarity of the initial dense Fe layers. The *z*-coordinate of the relaxed configuration is labeled in **Figure S4b** to accentuate this finding. The three unique Fe atoms in the dense layers have three unique *z*-coordinate, which is indicative of the lack of co-planar arrangement. The co-planar arrangement of the dense layers is an important fingerprint of the high-symmetry B20 materials class (see **Figure S1**). Intriguingly, this atomic arrangement also has a profound impact on the sparse and dense Ge layers. The Ge atoms also lose the coplanar atomic arrangement due to this configuration. It is important to note that such a disruption is not observed in **Figure S3**.

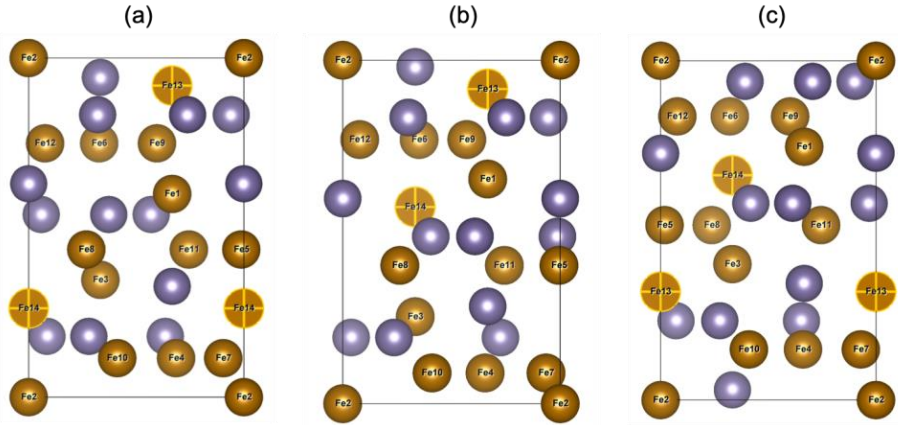

**Figure S3**. Crystal structures of fully relaxed $Fe_{1.17}Ge$ composition with two excess Fe atoms in the sparse layers (labeled as "Fe13" and "Fe14"). All excess Fe atoms form their own "new" sparse layers in the structure.

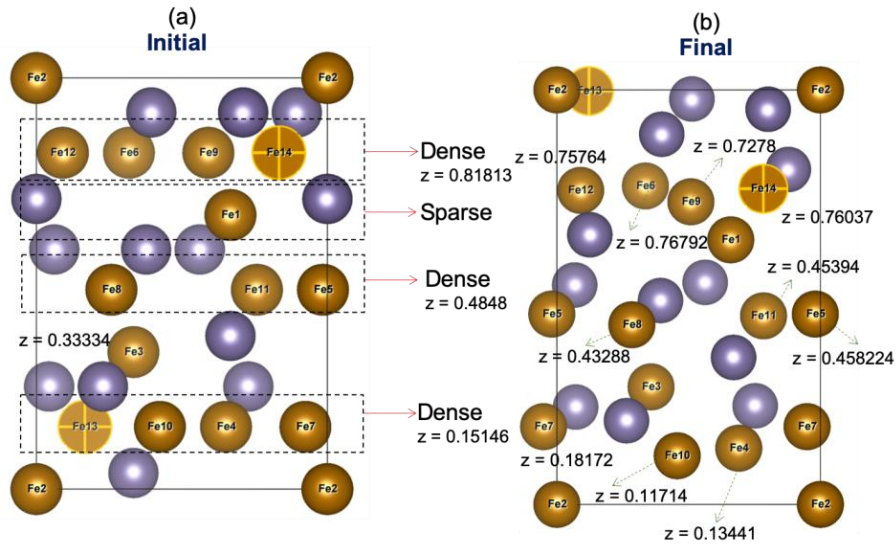

**Figure S4**. Crystal structures of (a) an initial and (b) fully relaxed $Fe_{1.17}Ge$ composition with two excess Fe atoms (labeled as "Fe13" and "Fe14"). In (a) both excess Fe atoms are in the dense layers. In (b) atom Fe13 is on the sparse layer and atom Fe14 disrupts the symmetry of the dense layer and is no longer coplanar with the initial dense layer.

We identify the structures shown in **Figure S3** to have a higher symmetry than that shown in **Figure S4b**. The relaxed final structure in **Figure S4b** has an out-of-plane lattice constant value of 8.8895 Å. However, the structure shown in **Figure S4b** is 58.4 meV/atom *lower* in energy compared to that shown in **Figure S3**. Our DFT calculations suggest that the lowest energy configuration for the $Fe_{1.17}Ge$ composition has a lower symmetry.

### $Fe_{1.25}Ge$ with three excess Fe atoms:

Finally, we also explored another atomic configuration with three excess Fe atoms in the sparse layers. In this simulation, we relaxed all internal coordinates and the out-of-plane lattice constant. The fully relaxed structure is shown in **Figure S5**, where the excess Fe atoms are labeled as "Fe13", "Fe14" and "Fe15". All excess Fe atoms occupy the sparse layers in the structure. Similar to the structures shown in **Figure S3**, the excess Fe atoms form their own *new* sparse layers in the crystal structure. Further, the coplanarity of the dense Fe and Ge layers are also maintained in this atomic arrangement (indicative of a crystal structure with higher symmetry). The calculated out-of-plane lattice constant is 9.5369 Å.

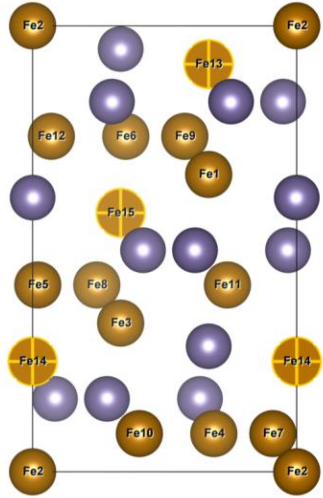

**Figure S5**. Crystal structure of a fully relaxed $Fe_{1.25}Ge$ composition with three excess Fe atoms (labeled as "Fe13", "Fe14", and "Fe15") in the sparse layers.

One of the main findings from the DFT calculations is that when the excess Fe atoms are located in the sparse layers, we get a high symmetry structure that is similar to those seen in the experimentally grown thin films. Although there are lower energy structures, these atomic arrangements occur at the expense of the crystal symmetry. Since the X-ray diffraction data and transmission electron micrograph on the experimentally grown thin films showed strong evidence for a high symmetry B20-like crystal structure, we conclude that the structures shown in **Figure S3**, where the excess Fe atoms occupy the sparse layers, are more representative of the symmetry of our thin films.

## Supercells of bulk FeGe simulations

In addition to the (111)-oriented FeGe crystal structures, we also performed DFT calculations on the bulk FeGe. We created a 2x2x1 supercell with 32 atoms (16 Fe and 16 Ge atoms). The crystal structure is shown in **Figure S6**.

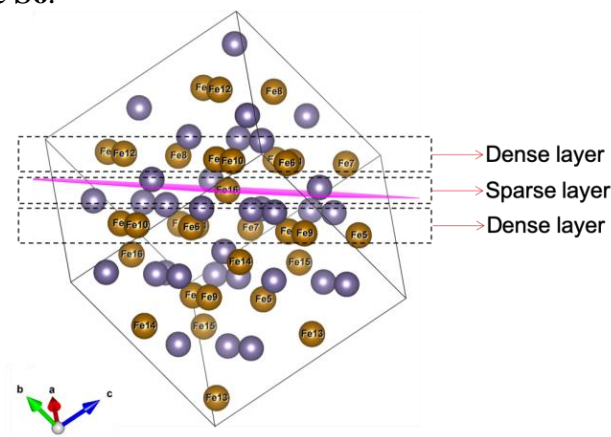

**Figure S6**. Crystal structure of bulk, stoichiometric FeGe. The crystal structure is rotated to highlight the atomic arrangements in dense and sparse layers. The magenta color plane is drawn as a guide to the eye, which intersects atom Fe16 on the sparse layer. Fe- and Ge-atoms are shown in brown and purple colors, respectively.

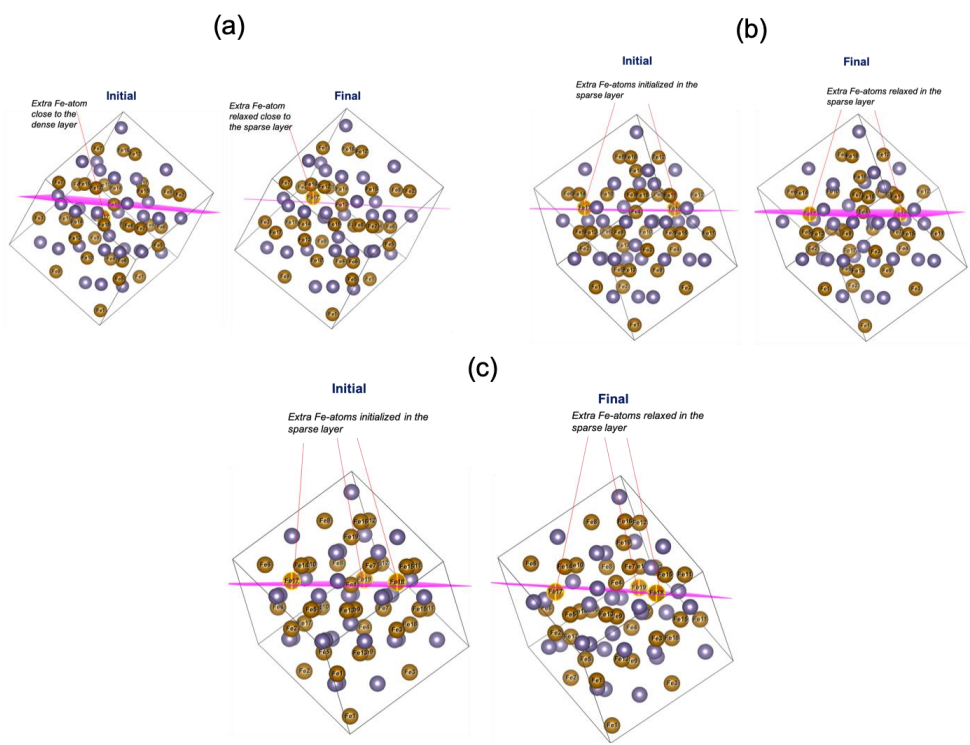

**Figure S7**. Crystal structures of bulk, FeGe supercells with (a) one excess Fe atom ("Fe17"), (b) two excess Fe atoms ("Fe18" and "Fe19"), and (c) three excess Fe atoms ("Fe17", "Fe18" and "Fe19").

We then systematically added one, two, and three excess Fe atoms into the structure and allowed the internal coordinates and out-of-plane lattice constant to relax. The results are shown in **Figure S7**. In **Figure S7a**, we initialized the structure with the excess Fe atom ("Fe17") in the dense layer. However, the Fe17 atom diffused out of the dense layer and relaxed in the near-by sparse layer. In **Figures S7b** and **S7c**, we show the results for the two and three excess Fe atoms, respectively. In both cases, we initialized the DFT calculations with the excess Fe atoms in the sparse layer. The calculations converged with those excess Fe atoms relaxing in the sparse layer. All DFT results paint a consistent picture indicating the preferential occupation of excess Fe atoms in the sparse layers of the B20 crystal structure.

## 2. MFM images without circles

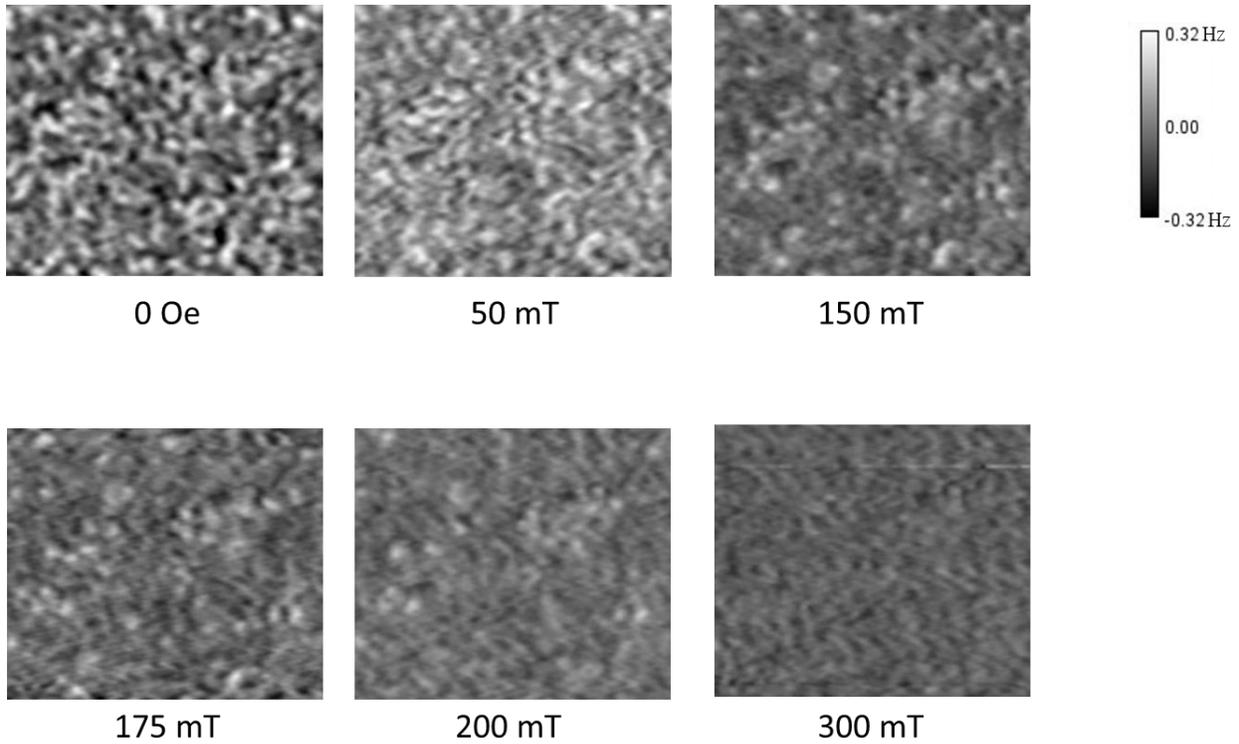

**Figure S8.** MFM images of Figure 3 of the main text without circling the skyrmions.

## 3. Topological Hall Effect Measurements and Analysis

The total Hall resistivity, $\rho_{xy}$, is the sum of three contributions: (1) the ordinary Hall effect $\rho_{OH}$, a result of the Lorentz force, that is proportional to the magnetic field strength ($H$); (2) the anomalous Hall effect $\rho_{AH}$, which is proportional to z-component of magnetization ($M_z$); and (3) the topological Hall effect $\rho_{TH}$, resulting from the reflecting of electrons or holes by real space topological spin textures, i.e.,

$$\rho_{xy} = \rho_{OH} + \rho_{AH} + \rho_{TH} = R_0 H + R_{AH} M + \rho_{TH} \qquad (1)$$

where $R_0$ and $R_{AH}$ are the ordinary Hall coefficient and anomalous Hall coefficient, respectively. Dividing both sides of equation (1) by $H$,

$$\rho_{xy}/H = R_0 + R_{AH} M/H + \rho_{TH}/H \qquad (2)$$

In the high field range, where magnetizations have been totally saturated and the topological spin textures have been eliminated, the $\rho_{AHE} = 0$, then equation (2) can be simplified to

$$\rho_{xy}/H = R_0 + R_{AH} M/H \qquad (3)$$

Therefore, we measure both the $\rho_{xy}$ vs. $H$ and $M$ vs. $H$ on the same sample. Plotting $\rho_{xy}/H$ vs. $M/H$ at the range of $H$ that is larger than the saturation field and doing linear fit, then we could obtain the $R_0$ and $R_{AH}$ from the y-intercept and the slope of the fitting line. Consequently, we could separate the three contributions from Hall resistivity.

Based on the above analysis, we obtained the phase diagram of $\rho_{THE}$ of our Fe-rich Fe$_{1.2}$Ge using the following procedure:

(1) Measure both the $\rho_{xy}$ vs. $H$ and $M$ vs. $H$ of a Fe-rich Fe$_{1.2}$Ge sample. These start at +7 Tesla and ramp down to -7 Tesla and is denoted as $\rho_{xy}^-(H)$, followed by a scan from -7 Tesla to +7 Tesla and is denoted as $\rho_{xy}^+(H)$. Considering the Hall resistivity should be antisymmetric in $H$, we remove any longitudinal resistance leaking into the transverse signal (which is symmetric in $H$) by anti-symmetrizing the resistivity data according to $\tilde{\rho}_{xy}^+(H) = \frac{1}{2}[\rho_{xy}^+(H) - \rho_{xy}^-(-H)]$ and $\tilde{\rho}_{xy}^-(H) = \frac{1}{2}[\rho_{xy}^-(H) - \rho_{xy}^+(-H)]$.
(2) Plot $\tilde{\rho}_{xy}^+/H$ vs. $M/H$ at the range of $H > 2$ T (in saturation).
(3) Linearly fit $\tilde{\rho}_{xy}^+/H$ vs. $M/H$ to obtain $R_0$ and $R_{AH}$ from the y-intercept and the slope.
(4) Calculate $\rho_{OH} = R_0 H$ and $\rho_{OH} = R_{AH} M$
(5) Calculate $\rho_{TH} = \tilde{\rho}_{xy}^+ - \rho_{OH} - \rho_{AH}$ (and $\rho_{TH} = \tilde{\rho}_{xy}^- - \rho_{OH} - \rho_{AH}$)
(6) Repeat steps (1)-(5) at different temperatures, we obtain the $\rho_{TH}$ at different temperatures to obtain a phase diagram. By convention, we plot $\rho_{TH} = \tilde{\rho}_{xy}^+ - \rho_{OH} - \rho_{AH}$ vs. $T$ and $H$ for the phase diagram.

**Figures S9 and S10** show a representative set of Hall resistivity data analysis results which obtain at a temperature of 340 K. As one can note that a clear topological Hall resistance component, with unique antisymmetric shape similar to that of previous reported ordinary FeGe films,[1] has been obtained even at a temperature much higher than the Curie temperature of ordinary FeGe.

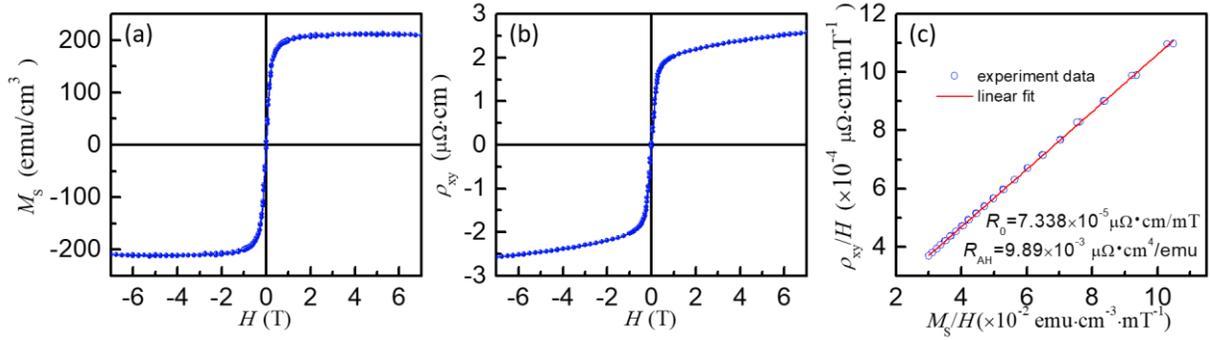

**Figure S9.** (a) Hysteresis loop and (b) Hall resistivity of a Fe-rich $Fe_{1.2}Ge$ sample, both of which are measured at 340 K in a Quantum Design PPMS using exactly the same field scanning procedure. (c) Linear fit of the $\rho_{xy}/H$ vs. $M/H$ at $H > 2$ T. The $\rho_{xy}$ have been anti-symmetrized.

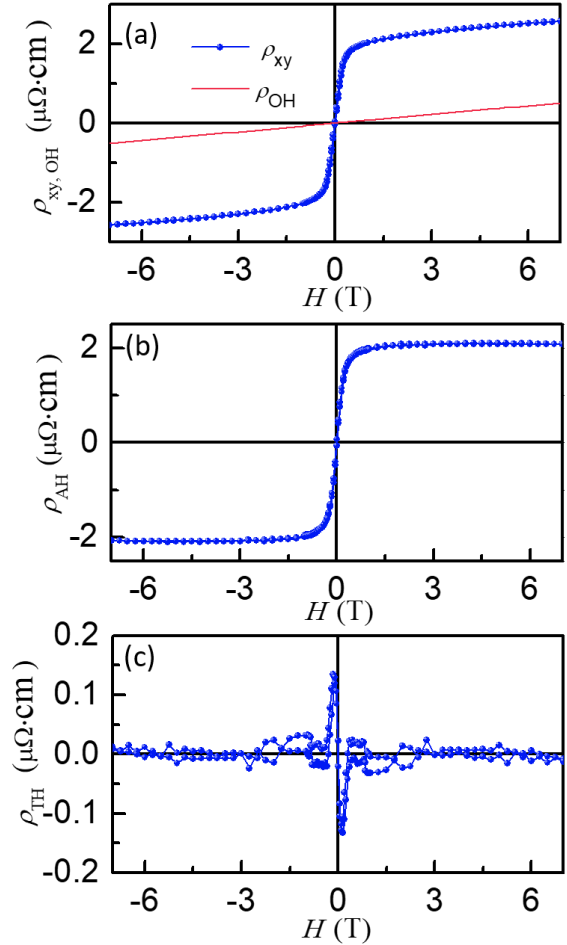

**Figure S10.** (a) $\rho_{xy}$ and $\rho_{OH}$, (b) $\rho_{AH}$, and (c) $\rho_{TH}$ of the Fe-rich $Fe_{1.2}Ge$ sample at 340 K, obtained using our Hall resistivity analysis method. The $\rho_{xy}$ have been anti-symmetrized (i.e. $\tilde{\rho}_{xy}^+$ and $\tilde{\rho}_{xy}^-$)

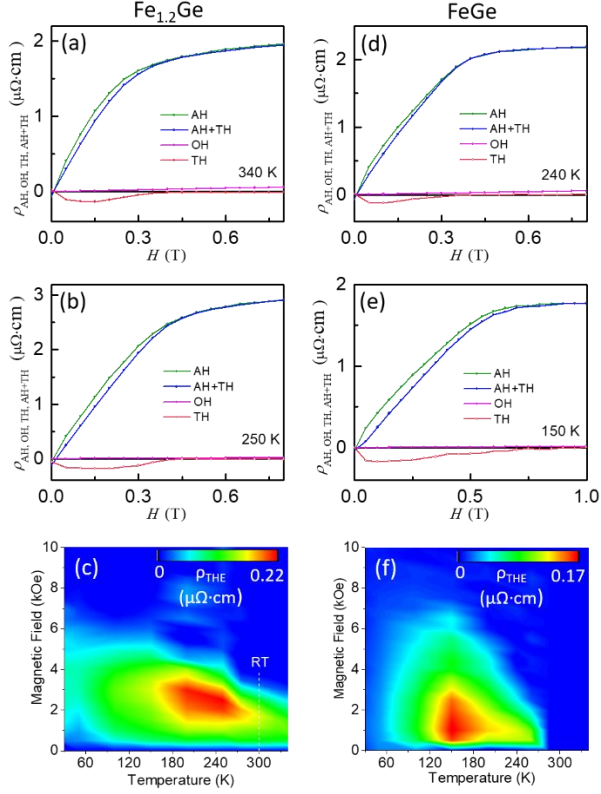

**Figure S11.** Comparing the topological resistivity of Fe-rich $Fe_{1.2}Ge$ and ordinary FeGe. The fourth quarter part of $\rho_{AH}$, $\rho_{AH}+\rho_{TH}$ and $\rho_{TH}$, which have the same magnetic field scanning procedure as that for taking MFM images, of (a)(b) $Fe_{1.2}Ge$, and (d)(e) FeGe at several representative temperatures. $\rho_{TH}$ vs. field and temperature phase diagraph of (c) $Fe_{1.2}Ge$ and (f) FeGe.

Note that the relative values of $\rho_{TH}$ to that of $\rho_{xy}$ and $\rho_{AH}$ are always very small. Therefore, the hysteresis loop and Hall resistivity measurements and data analysis procedures need to be carried out very carefully. To minimize the contamination of $\rho_{TH}$ by any artificial effect, the following schemes had been taken: (1) In previously reported topological Hall resistivity measurements,[1] the hysteresis loops and Hall resistivity loops were always measured using two facilities. For example, the hysteresis loops were measured by SQUID and the Hall resistivity loops were measured by PPMS. To make sure the magnetic fields in corresponding magnetic and transport data are exactly the same, both measurements were both performed in the same Quantum Design PPMS system and the employed field scan procedures were identical. (2) The longitudinal resistivity was also measured during the Hall resistivity measurements, so that the longitudinal resistivity mixed into Hall resistivity due to device asymmetry can be well subtracted. We found that the most reliable method for removing the longitudinal resistivity is the anti-symmetrization procedure discussed earlier.

In order to validate of our approach for obtaining $\rho_{TH}$, we have also fabricated an ordinary FeGe film, in which we had observed skyrmions with size similar to previous reports using L-TEM.[2,3] The obtained topological resistivity of Fe-rich $Fe_{1.2}Ge$ and ordinary FeGe films are presented in **Figure S11**.

## 4. Topological Hall Resistivity (THR) Estimates

We use the Drude model-like analysis of P. Bruno et al.[4] to make a simple estimate of the topological Hall effect. We could use Eqs. (13,14) of N. Verma et al.[5] for an arbitrary band structure, but our goal here is just an order-of-magnitude estimate.

We make a further simplification by letting the scattering time $t_s = t$ independent of the spin label s. Then eq. (6) and (7) of Bruno PRL lead to a THR result which can be written as

$$\rho_{xy} = P\, R_H\, B_{eff}$$

where $R_H = 1/ne$ is the usual Hall coefficient, $B_{eff} = n_{sk}(h/e)$ = (skyrmion density x flux quantum) is the "effective magnetic field", and $P = (n_u - n_d)/(n_u + n_d)$ is conduction electron polarization with u = up and d = down. Going beyond a Drude-like theory, P will actually involve the conduction electron DOS at the Fermi surface, i.e., $n_u$ will be replaced by $N_u(0)$ and $n_d$ by $N_d(0)$; but this will not affect the simple estimates. P should lie between 0 and 1, and for now we will just treat it as an unknown of order 1.

Using the measured $R_H$ in the ordinary Hall regime at high fields, and $h/e = 4.14 \times 10^{-15}$ Tm$^2$, we get the following results

|  | FeGe (200K, 2000 Oe) | Fe-rich FeGe (300K, 1500 Oe) |
|---|---|---|
| Measured $R_H$ (µΩ cm/T) | 0.39 | 0.61 |
| Measured $n_{sk}$ (m$^{-2}$) | 60.7 x 10$^{12}$ | 25.5 x 10$^{12}$ |
| Estimated $B_{eff}$ (T) | 0.25 | 0.11 |
| Estimated topo $\rho_{xy}$ (µΩ cm) | 0.098 | 0.067 |
| Measured topo $\rho_{xy}$ (µΩ cm) | 0.1528 | 0.1548 |

We see that this simple theory is off by a factor of 2-3 from the THR Data. We assumed P = 1 so, if P is actually 0.1 we are further off by a factor of 10. The sign is not issue since P can be either positive or negative.

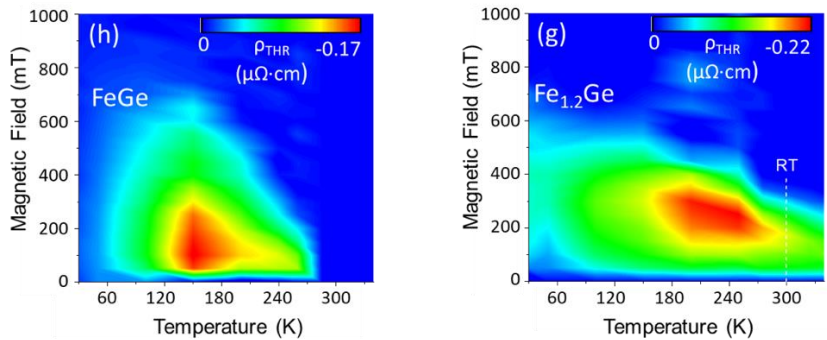

**Figure S12.** Plots of topological Hall resistivity as a function of magnetic field and temperature for an epitaxial FeGe film (left) and Fe-rich FeGe film (right).

## 5. Determination of Micromagnetic Parameters

We discuss the determination of the micromagnetic parameters: saturation magnetization $M_s$, exchange stiffness $A_{ex}$, bulk Dzyaloshinskii-Moriya interaction (DMI) $D$, magnetic anisotropy $K_u$, and the Gilbert damping $\alpha$. These parameters are a succinct way to describe the magnetism in a material via the effective free energy function

$$\mathcal{F} = A_{ex}(\nabla \boldsymbol{m})^2 + D\boldsymbol{m} \cdot (\nabla \times \boldsymbol{m}) - K_u m_z^2 - H m_z$$

Here $\boldsymbol{m}$ is the unit vector of magnetization ( $\boldsymbol{m} = \boldsymbol{M}/M_S$ ), and the $D$ term describes the bulk DMI characteristic of the non-centrosymmetric B20 crystal structure. The damping $\alpha$ enters the Landau Lifshitz Gilbert (LLG) equation.

One of the key questions in skyrmion materials is: what parameters determine the size and stability of skyrmions? The simplest answer is that the size of skyrmions in determined by the ratio of the exchange to DMI, namely by the length scale $L_D = 4\pi A_{ex}/D$ [6,7]. However, there are several theoretical studies, especially of isolated skyrmions, which conclude that their size is *not* determined by $L_D$, and that it depends in an essential way on $K_u$ and $M_s$ [8–10].

In view of this, it is especially important to understand the parameters that characterize small, room-temperature skyrmions in Fe-rich FeGe. We will show below, combining micromagnetic simulations with our experimental measurements (magnetization, FMR, MFM and LTEM), that the size of Bloch skyrmions is indeed determined by $L_D \sim A_{ex}/D$ and weakly dependent of $K_u$ and $M_s$.

We start by summarize our results for room temperature values of the micromagnetic parameters in the table below.

| | |
|---|---|
| $M_s$ | 253 kA/m |
| $A_{ex}$ | 0.69 ± 0.07 pJ/m |
| $D$ | 0.25 mJ/m² |
| $K_u$ | 11.5 - 23 kJ/m³ |
| $\alpha$ | 0.019 |

The saturation magnetization $M_s$ is determined from the SQUID magnetometry measurement of *M vs. H* shown in **Figure 13** (the out-of-plane geometry is utilized to minimize effects of the point dipole approximation). We find that $M_s = 253$ kA/m, which is equivalent to 253 emu/cm³.

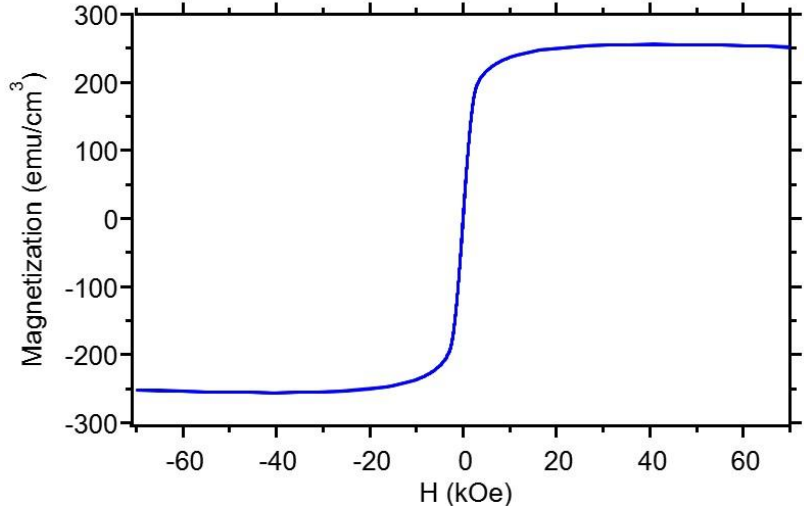

**Figure S13.** Out-of-plane hysteresis loop of 108 nm thick $Fe_{1.2}Ge$ film, with diamagnetic background subtracted, obtained at 300 K via SQUID magnetometry.

We use the following procedure to estimate the exchange $A_{ex}$. We start with the relation between the exchange stiffness and saturation magnetization at two different temperatures $T_1$ and $T_2$

$$\frac{A_{ex}(T_1)}{A_{ex}(T_2)} = \left(\frac{M_s(T_1)}{M_s(T_2)}\right)^2.$$

This can be understood by looking at how the microscopic exchange energy $J$ in a spin model is related to the stiffness $A_{ex}$ via $A_{ex} = J\, M_s^2\, a/2$, where $a$ is the lattice spacing. The above scaling follows from the fact that the microscopic exchange $J$ is a $T$-independent. Next, we use the fact that the transition temperatures $T_c$ of two magnetic materials A and B are both proportional to their $T=0$ exchange stiffnesses, so that

$$\frac{A_{ex,A}(0)}{A_{ex,B}(0)} = \frac{T_{c,A}}{T_{c,B}}.$$

Combining these two equations we find

$$A_{ex,A}(T_0) = A_{ex,B}(0) \frac{T_{c,A}}{T_{c,B}} \left(\frac{M_{s,A}(T_0)}{M_{s,A}(0)}\right)^2.$$

We choose $T_0 = 300$ K (room temperature) for the system A of interest, namely, Fe-rich FeGe and as our reference we choose a closely related material B = $Mn_{0.08}Fe_{0.92}Ge$ for which the required data $A_{ex,B}(0) = 1.47 \pm 0.14$ pJ/m and $T_{c,B} = 260$ K are available from the literature[11]. From our own measurements on Fe-rich FeGe (= A) we have $M_{s,A}(300K) = 253$ kA/m from **Figure S13** and $M_{s,A}(0) = 470$ kA/m and $T_{c,A} = 420$ K from **Figure 2e** (main text). Using these inputs, we find that the room temperature exchange stiffness for Fe-rich FeGe is $A_{ex,A}(300K) = 0.69 \pm 0.07$ pJ/m.

The magnetic anisotropy and damping are determined from room temperature Ferromagnetic Resonance (FMR) experiments described below. The Gilbert damping $\alpha$ of Fe$_{1.2}$Ge is found to be $\alpha = 0.019$. We can constrain $K_u$ to lie in the range 11.5 - 23 kJ/m$^3$ (or $1.15 \times 10^5 - 2.3 \times 10^5$ erg/cm$^3$). Note that the positive sign indicates an easy axis (out-of-plane) anisotropy for our Fe-rich FeGe thin films. We will show below that the uncertainty in our $K_u$ estimates does *not* impact our micromagnetic simulations of skyrmion size and stability.

Absent a direct measurement of the Dzyaloshinskii-Moriya interaction, we use micromagnetic simulations with a single free parameter $D$ to fit the skyrmion size, R$_{sk}$ = 10.7 ± 4.7 nm (HWHM), measured by our LTEM and MFM experiments described in the paper. The micromagnetic results were independently checked using MuMax3[12] and our custom simulation software. We find that the skyrmion size depends crucially on DMI, and 11 nm radius skyrmions are obtained for $D = 0.25$ mJ/m$^2$ as shown in **Figure S14**. These results are only weakly dependent on the value of $K_u$, as seen from **Figure S15**, and also of $M_s$, as seen from **Figure S16**.

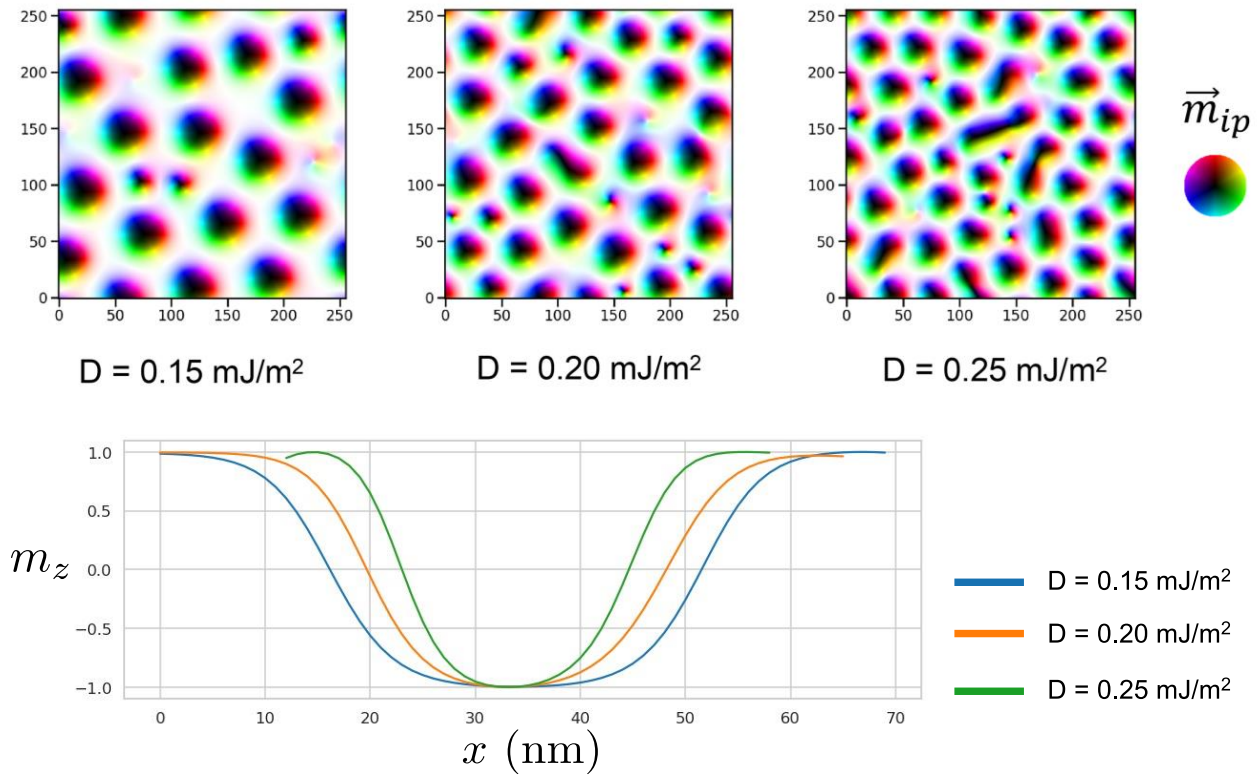

**Figure S14:** Results of micromagnetic simulations for skyrmion with the experimentally measured $M_s$ = 253 kA/m, $A_{ex}$ = 0.69 pJ/m, and $K_u$ = 17 kJ/m$^3$ for a 108 nm thick film in an external field of H = 0.16 T. We show results for three different DMI values. Choosing $D = 0.25$ mJ/m$^2$ leads to skyrmions of 22 nm diameter, consistent. with our LTEM and MFM experiments. The upper panels show the in-plane (ip) magnetization, and the lower panel shows line cuts of the spatial variation of m$_z$, from which we define the skyrmion diameter using the zero crossings.

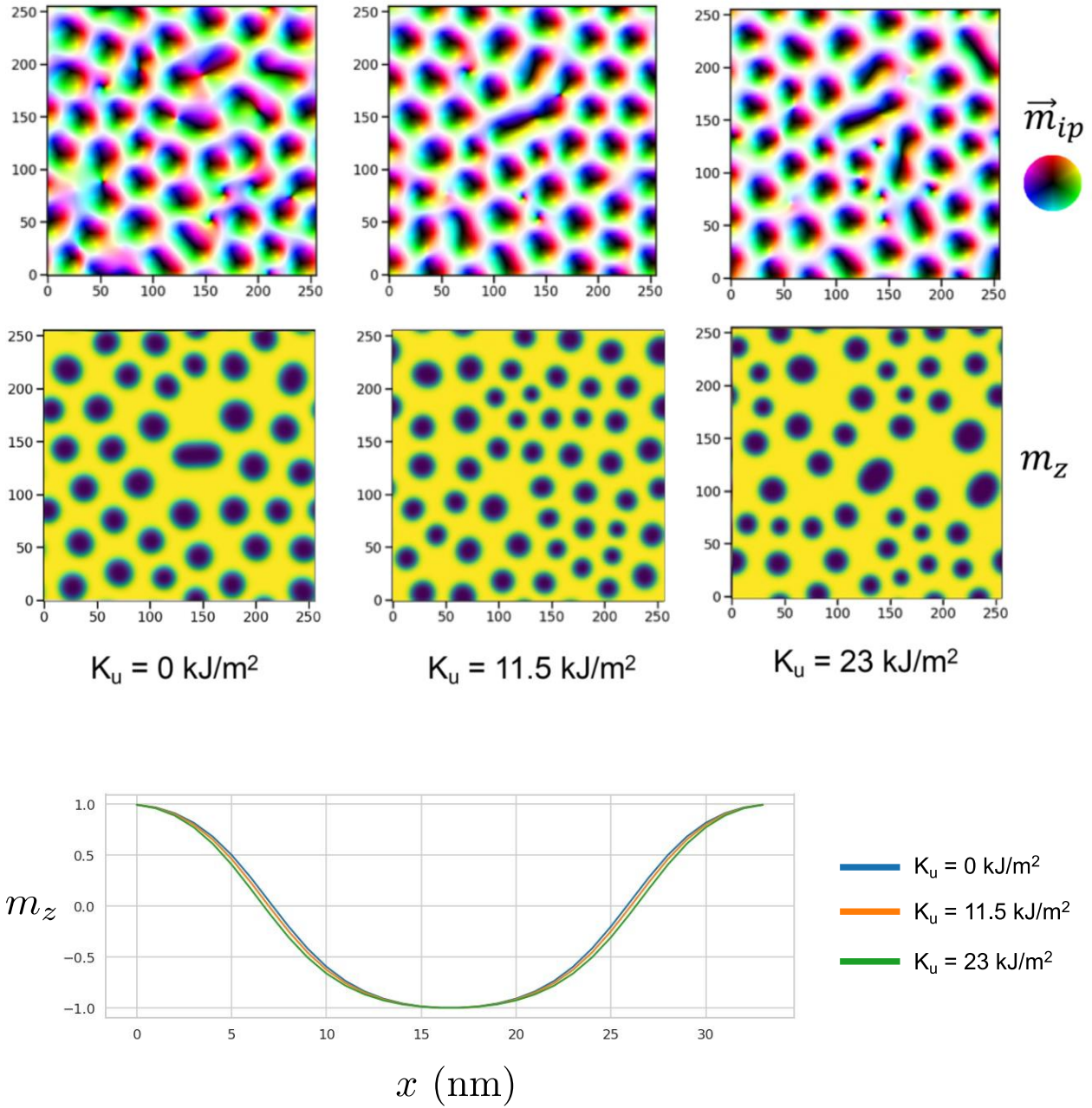

**Figure S15:** Results of micromagnetic simulations for skyrmion with the experimentally measured $M_s$ = 253 kA/m, $A_{ex}$ = 0.69 pJ/m, for a 108 nm thick film in an external field of H = 0.16 T. We chose a DMI $D$ = 0.25 mJ/m$^2$ in order to get a skyrmion radius that matches our LTEM and MFM experiments. We show that the results are substantially independent of the magnetic anisotropy $K_u$. Note that our FMR measurements lead to $K_u$ estimates in the range 11.5 - 23 kJ/m$^3$. The upper panels show the in-plane magnetization, the middle panels show m$_z$, while the lower panel shows line cuts of the spatial variation of m$_z$, from which we define the skyrmion diameter using the zero crossings.

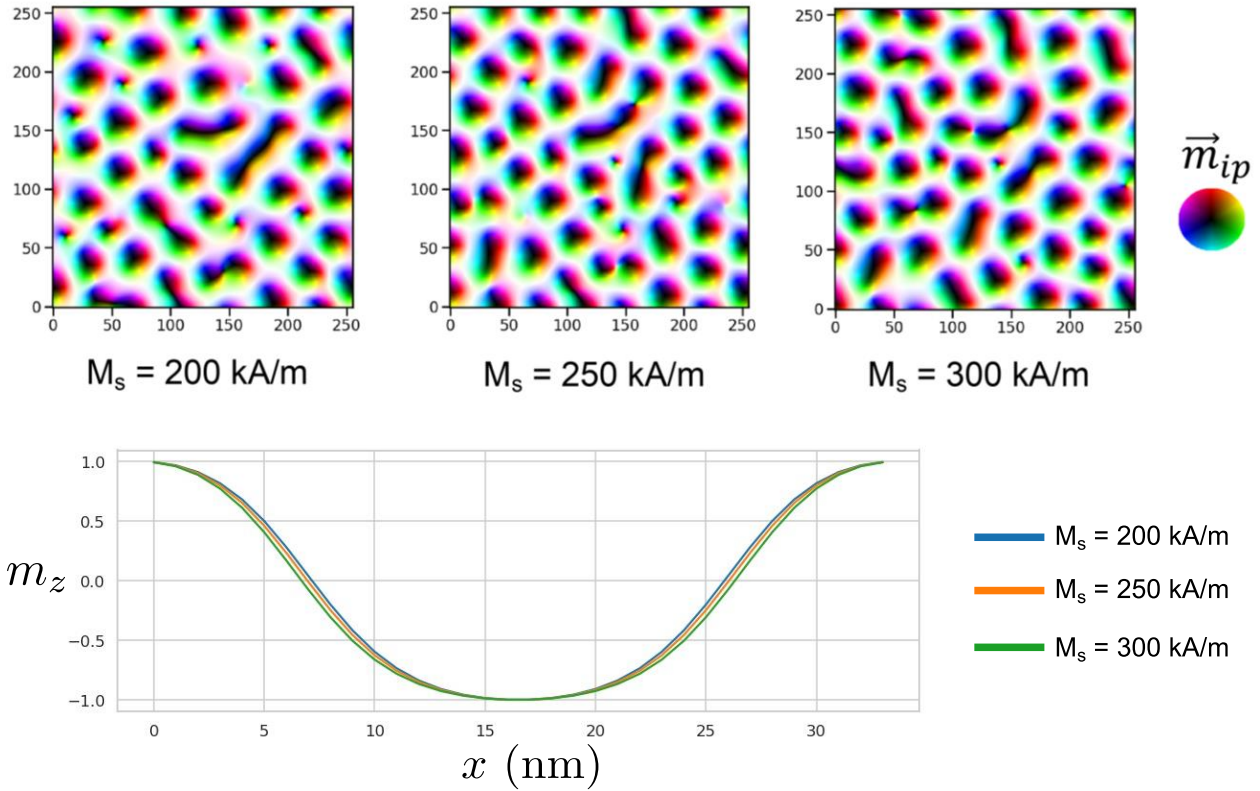

**Figure S16:** Results of micromagnetic simulations for skyrmion with the experimentally measured $A_{ex} = 0.69$ pJ/m, for a 108 nm thick film in an external field of H = 0.16 T. We chose a DMI $D = 0.25$ mJ/m$^2$ and $K_u = 17$ kJ/m$^3$. The upper panels show the in-plane magnetization, while the lower panel shows the spatial variation of m$_z$, from which we define the skyrmion diameter using the zero crossings. The results for skyrmion size are substantially independent over a range of $M_s$ values around the experimentally measured $M_s = 253$ kA/m.

### Ferromagnetic Resonance measurements

We conducted broad band Ferromagnetic Resonance (FMR) studies of 54 nm thick Fe$_{1.2}$Ge film grown on 5 nm FeGe layer on Si substrate at room temperature and under ambient conditions. The external magnetic field was applied in the plane of the film.

### Magnetic anisotropy

A series of FMR spectra, acquired via magnetic field modulation, at several frequencies of applied rf radiation is shown in **Figure S17a**. The resonant field of the signal $H_{res}$ and the full width at half-maximum linewidth $\Delta H$ at each rf frequency was determined using Lorentzian lineshape fitting function incorporating symmetric and anti-symmetric contributions:

$$V = \frac{S + A(H - H_{res})}{((H - H_{res})^2 + (\Delta H/2)^2)^2}$$

Here $V$ is the amplitude of the field modulated FMR signal at a given static field $H$, and $S$ and $A$ are the coefficients describing the symmetric and the anti-symmetric parts of the Lorentzian lineshape, respectively. The resulting dispersion relationship and the frequency dependence of the linewidth $\Delta H$ are shown in **Figure S17b** and **Figure S17c**, respectively.

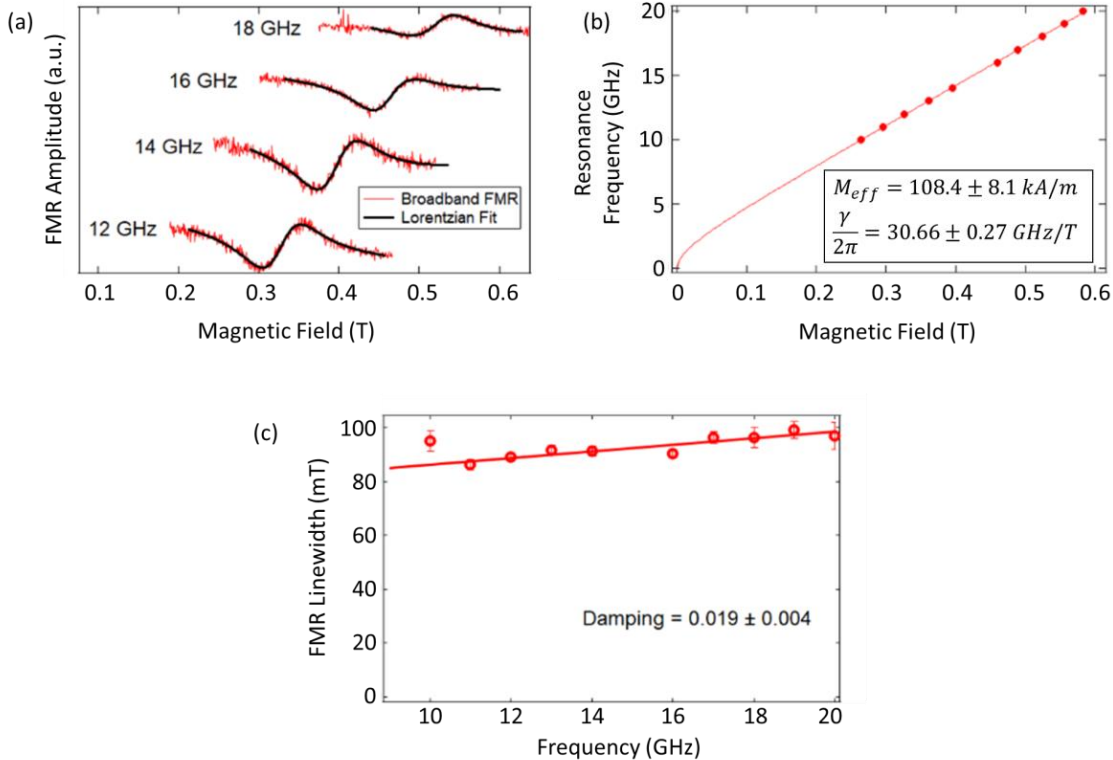

**Figure S17.** Room temperature broadband FMR of $Fe_{1.2}Ge$ thin film. (a) Broadband field-modulated FMR response of $Fe_{1.2}Ge$ with fits to a combined symmetric and antisymmetric Lorentzian lineshape. (b) Fit of the FMR resonance condition fit to the in-plane Kittel response for a ferromagnetic thin film. (c) Linear fit of the FMR linewidth versus frequency to extract the Gilbert damping parameter α.

We analyze the obtained dispersion relationship using the Kittel equation for the FMR resonance condition for a thin ferromagnetic film with the static field applied in the film plane:

$$f_{res} = \frac{\gamma \mu_0}{2\pi} \sqrt{H(H + M_{eff})}$$

where $f_{res}$ is the resonance frequency, $H$ is the applied magnetic field, $\gamma$ is the gyromagnetic ratio, and $M_{eff}$ is the effective magnetization. The fit of the experimental data results in the effective magnetization of $M_{eff} = 108.4$ kA/m ($4\pi M_{eff} = 1362.1$ Gauss). Usually, the obtained value of $M_{eff}$ is used to determine the uniaxial magnetocrystalline anisotropy energy density $K_u$ following the expression

$$K_u = \frac{\mu_0 M_s}{2}(M_s - M_{eff}),$$

if the saturation magnetization $M_S$ is known. Interpretation of the FMR data this way is complicated by the fact that the FMR measurements were conducted in the magnetic field range of approximately 0.3 – 0.6 T where, according to the magnetometry data shown in **Figure S18**, the sample is not fully saturated. As a result, the value of the saturation magnetization is between approximately 200 kA/m and the value of 253 kA/m, the latter determined in the fully saturated sample. As a result of such an ambiguity in determination of $M_S$, only a range of values of $K_u$ can be determined, which we estimate to be 11.5 – 23.0 kJ/m³, thus indicating that the magnetocrystalline anisotropy favors the out-of-plane axis. Using micromagnetic simulations we show in **Figure S15** that the uncertainty in the value of $K_u$ does not play an important role in determining the skyrmion size and stability.

These results are consistent with the in-plane and out-of-plane SQUID loops (**Figure S18**), which show an overall preference for in-plane magnetization as indicated by the lower saturation field for the in-plane loop. While the positive $K_u$ favors out-of-plane magnetization, the magnetic shape anisotropy $\left(\frac{1}{2}\mu_0 M_S^2\right)$ favoring in-plane magnetization is stronger, leading to an overall preference for in-plane magnetization.

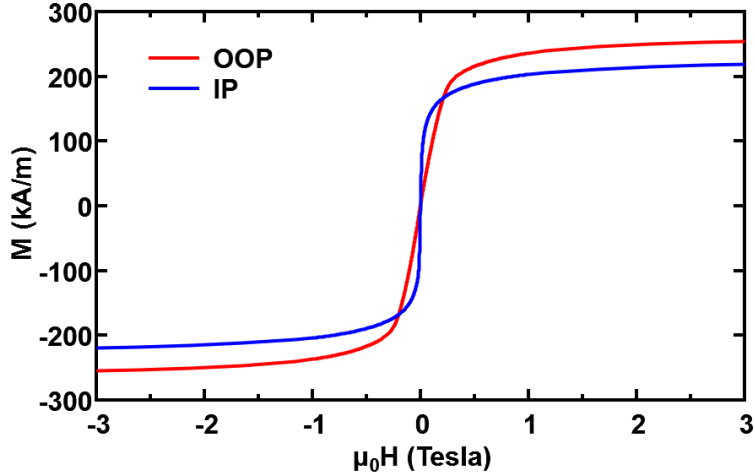

**Figure S18.** Out-of-plane (OOP, red) and in-plane (IP, blue) hysteresis loops of Fe$_{1.2}$Ge obtained at 300 K.

## Gilbert damping

The results of FMR measurements shown in **Figure S17c** were used to determine Gilbert damping parameter $\alpha$ of Fe$_{1.2}$Ge. This was extracted by fitting the frequency dependence of the linewidth $\Delta H$ with the expression:[13]

$$\Delta H = \Delta H_0 + \frac{4\pi\alpha f}{\mu_0 \gamma}$$

where $f$ is the rf frequency, $\Delta H_0$ the inhomogeneous broadening contribution to the linewidth, and $\gamma$ is the gyromagnetic ratio. The fit of the frequency dependence of the linewidth results in $\alpha = 0.019$.

## References

1. Ahmed, A. S., Rowland, J., Esser, B. D., Dunsiger, S. R., McComb, D. W., Randeria, M. & Kawakami, R. K. Chiral bobbers and skyrmions in epitaxial FeGe/Si(111) films. *Phys. Rev. Materials* **2,** 041401(R) (2018).

2. Bagues, N., Wang, B., Liu, T., Selcu, C., Boona, S., Kawakami, R., Randeria, M. & McComb, D. Imaging of Magnetic Textures in Polycrystalline FeGe Thin Films via in-situ Lorentz Transmission Electron Microscopy. *Microscopy and Microanalysis* **26,** 1700–1702 (2020).

3. Wang, B., Bagués, N., Liu, T., Kawakami, R. K. & McComb, D. W. Extracting weak magnetic contrast from complex background contrast in plan-view FeGe thin films. *Ultramicroscopy* **232,** 113395 (2022).

4. Bruno, P., Dugaev, V. K. & Taillefumier, M. Topological Hall Effect and Berry Phase in Magnetic Nanostructures. *Phys. Rev. Lett.* **93,** 096806 (2004).

5. Verma, N., Addison, Z. & Randeria, M. Unified theory of the anomalous and topological Hall effects with phase-space Berry curvatures. *Science Advances* **8,** eabq2765 (2022).

6. Nagaosa, N. & Tokura, Y. Topological properties and dynamics of magnetic skyrmions. *Nature Nanotechnology* **8,** 899–911 (2013).

7. Banerjee, S., Rowland, J., Erten, O. & Randeria, M. Enhanced Stability of Skyrmions in Two-Dimensional Chiral Magnets with Rashba Spin-Orbit Coupling. *Phys. Rev. X* **4,** 031045 (2014).

8. Büttner, F., Lemesh, I. & Beach, G. S. D. Theory of isolated magnetic skyrmions: From fundamentals to room temperature applications. *Sci. Rep.* **8,** 4464 (2018).

9. Wang, X. S., Yuan, H. Y. & Wang, X. R. A theory on skyrmion size. *Commun. Phys.* **1,** 1–7 (2018).

10. Bera, S. & Mandal, S. S. Theory of the skyrmion, meron, antiskyrmion, and antimeron in chiral magnets. *Phys. Rev. Research* **1,** 033109 (2019).

11. Turgut, E., Paik, H., Nguyen, K., Muller, D. A., Schlom, D. G. & Fuchs, G. D. Engineering Dzyaloshinskii-Moriya interaction in B20 thin-film chiral magnets. *Phys. Rev. Materials* **2,** 074404 (2018).

12. Vansteenkiste, A., Leliaert, J., Dvornik, M., Helsen, M., Garcia-Sanchez, F. & Van Waeyenberge, B. The design and verification of MuMax3. *AIP Advances* **4,** 107133 (2014).

13. Kalarickal, S. S., Krivosik, P., Wu, M., Patton, C. E., Schneider, M. L., Kabos, P., Silva, T. J. & Nibarger, J. P. Ferromagnetic resonance linewidth in metallic thin films: Comparison of measurement methods. *Journal of Applied Physics* **99,** 093909 (2006).